\documentclass[a4paper,10pt]{article}
\usepackage[top=1in,bottom=1in,left=0.75in,right=0.75in]{geometry}
\usepackage [dvips]{graphicx}
\usepackage{amssymb,color,amsmath,fancyhdr,setspace,caption}

 \title{Phase coexistence phenomena in an extreme case\\ of the misanthrope process with open boundaries}
 \def\shorttitle{Misanthrope process with open boundaries}
 \author{Chikashi Arita$^1$, Chihiro Matsui$^2$} 
\def\shortauthor{C Arita, C Matsui} 
 \def\address{1 Theoretische Physik, Universit\"at des Saarlandes, 66041 Saarbr\"ucken, Germany\\
   2 Mathematical Informatics, The University of Tokyo, 113-8656 Tokyo, Japan}
 \def\abst{The misanthrope process is a class of stochastic interacting particle systems, generalizing the simple exclusion process. It allows each site of the lattice to accommodate more than one particle. We consider a special case of the one dimensional misanthrope process whose probability distribution is completely equivalent to the ordinary simple exclusion process under the periodic boundary condition. By imposing open boundaries, high- and low-density domains can coexist in the system, which we investigate by Monte Carlo simulations. We examine finite-size corrections of density profiles and correlation functions, when the jump rule for particles is symmetric. Moreover, we study properties of delocalized and localized shocks in the case of the totally asymmetric jump rule. The localized shock slowly moves to its stable position in the bulk.}
 \date{\today}
\makeatletter
\def\@maketitle{ 
\begin{center} 
 \let \footnote \thanks
 {\LARGE\linespread{1.2}\selectfont\textbf{\@title}\par} \vskip 10mm 
 {\Large \@author} \vskip 5mm 
 {\address} \vskip 5mm 
 \textbf{Abstract} \end{center}
 \begin{quote} \abst \end{quote} \vskip 5mm 
\noindent\makebox[\linewidth]{\rule{\textwidth}{0.5pt}}}
\makeatother

\fancypagestyle{titlepage}{
 \lhead{} \chead{} \rhead{}
 \lfoot{} \cfoot{\thepage} \rfoot{} }

\begin{document}

\maketitle 

\thispagestyle{titlepage}

\section{Introduction}

Lattice gases serve as useful tools to model complex systems from molecular biology to vehicle traffic \cite{bib:SCN}. One of typical examples is the simple exclusion process (SEP), where each particle stochastically hops to one of its nearest-neighbor sites, if this target site is empty. As a paradigm of non-equilibrium statistical physics, the SEP has been intensively studied by exact solutions as well as phenomenological arguments \cite{bib:D,bib:BE,bib:DEHP,bib:SD}. In contrast to the ``simple'' one, the generalized exclusion processes allow each site to accommodate more than one particle \cite{bib:KLO,bib:TS,bib:AKM,bib:M}. The misanthrope process \cite{bib:C-T} is a class of generalized exclusion processes, which is relevant to modeling of traffic flows\cite{bib:Kan,bib:ERT}. In the infinite lattice or under the periodic boundary condition, its stationary state is given by the product of single-site weights and the relationship between the density and current (the so-called fundamental diagram) is exactly calculated, including the SEP as the simplest case. One can use these single-site weights to define the rates of particle injections and extractions at the boundaries of a finite chain (a so-called open system). Depending on these rates, the stationary current and density profile exhibit phase transitions. The phase diagram is phenomenologically determined by using the fundamental diagram, where the motion of a shock plays a key role \cite{bib:PS,bib:HKPS}. Therefore exploring properties of shocks is one of the most important subjects in lattice gasses.
 
 In this work, we study the misanthrope process with maximum occupancy number $ k=2 $. In the next section, we introduce notations for the misanthrope process with general $k$, and review some fundamental properties. We also explain our \textit{extreme} case, where we set the jump rates from $11$ to $20$ and $02$ to be 0. In the separated section ``Open boundary conditions'', we explain a general setup of the injection and extraction rates of particles. We briefly have a look at some formulas in the symmetric SEP, and give some remarks for our extreme misanthrope process. In ``Symmetric case'' (i.e. the same rates for the leftward and rightward jumps), we demonstrate a derivation of the diffusivity for the symmetric misanthrope process with $k=2$. For the extreme case, the density profile predicted in the limit of large system size is piecewise linear, which may be regarded as a second-order phase transition. We investigate the finite-size scaling of the density profile near this point, which is important from a perspective of statistical physics \cite{bib:HHL}. We examine correlation functions as well. In ``Totally asymmetric case'' (i.e. prohibiting leftward jumps) we explore properties of the shock, by introducing a microscopic definition of its position. There is a region in the phase diagram, where the shock moves to a stable point in the bulk \cite{bib:Kru}. We quantitatively show that the shock motion is very slow depending on the system size. 
 We describe main conclusions in the last section.

\section{Misanthrope process} 

Let us consider lattice gasses in one dimension, where each site can accommodate more than one but at most $k\in\mathbb N$ particles. (The case $k=\infty$ also can be considered.) The jump rates of particles depend on the occupation numbers of both departure and target sites: with the convention $ p_{mn} = 0 $ for $ n\ge k $, 
\begin{align}
 & \cdots\ m \, n\ \cdots \ \to\ \cdots \ m\!-\!1 \, n\!+\!1 \ \cdots \quad (\text{rate}\ p_{mn}), \\
 & \cdots\ m \, n\ \cdots \ \to\ \cdots \ m\!+\!1 \, n\!-\!1 \ \cdots \quad (\text{rate}\ q p_{nm}) .
\end{align} 
The misanthrope process is a class of generalized exclusion processes with some conditions on the jump rates, such that the stationary probability is given by a product of single-site weights \cite{bib:C-T}. 
 Thanks to the product measure, the rightward, leftward and \textit{total} currents are given as 
\begin{align} 
 \label{eq:Jto}
 J_\to = \sum_{ m \ge 1 \atop n \ge 0 } p_{mn} X_m X_n ,\ 
 J_\gets = q J_\to,\ 
 J = (1-q) J_\to\, , 
\end{align}
with $ X_n $ the probability of finding $n$ particles at each site. 

The case $ k=1 $ is the SEP. No condition on the jump rates is imposed, and we have simply $ X_0 = 1-\rho, X_1 = \rho $ and $ J = (1-q)p_{10}\rho(1-\rho) $ with the global density $ \rho $. 

\begin{figure} 
\vspace{-3mm}\begin{center}
  \includegraphics[width=40mm]{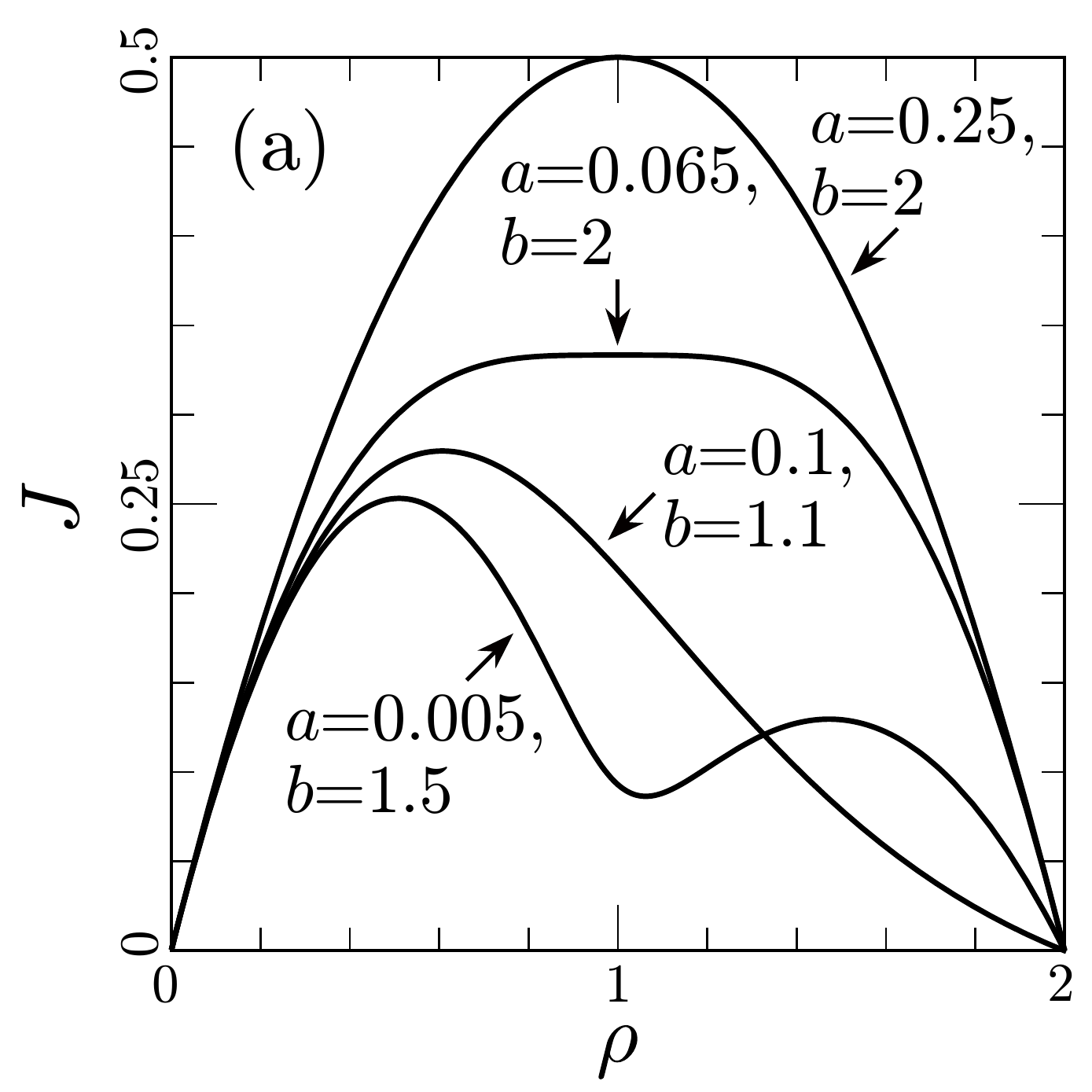}
  \includegraphics[width=40mm]{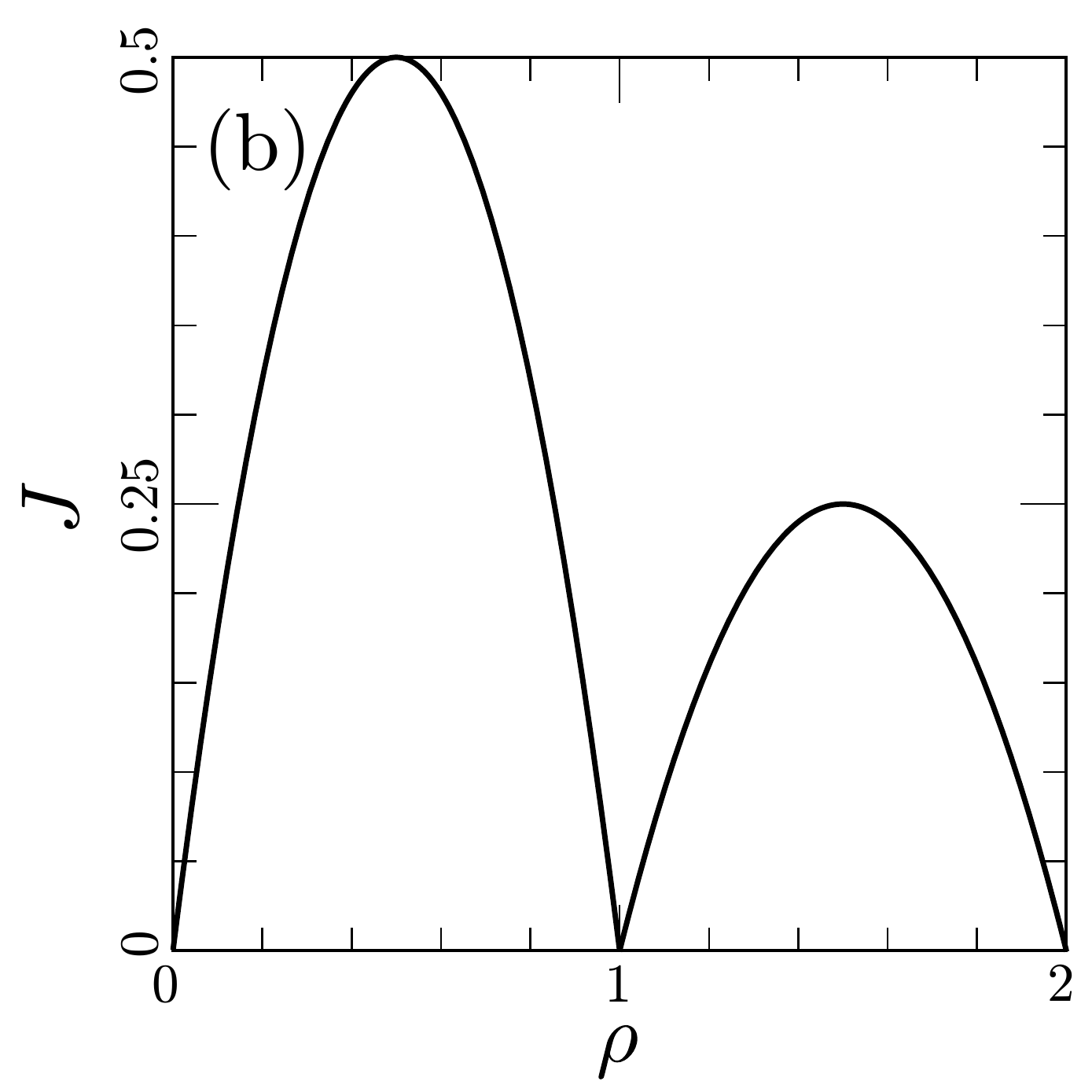}
 \caption{Fundamental diagrams of the misanthrope process for
 (a) various values of 
 $ a $ and $ b $ with $ (1-q)p_{10} = 1 $,
 and (b) the extreme case $ a= 0 $ with $(1-q)p_{10}=2 $ and $(1-q)p_{21}=1$. 
 \label{fig:fundamental-diagrams}} 
\end{center}
\end{figure}

For $k=2$, the following relation is imposed \cite{bib:TS,bib:C-T}:
\begin{align} \label{eq:r20=r21+r10} 
 p_{20} = p_{21} + p_{10} . 
\end{align}
 For convenience we also introduce notations $a$ and $b$ via 
\begin{align}
 a = p_{11} / p_{20} , \ b= p_{20} / p_{10}. 
\end{align}
The single-site weights for $k=2$ are given as \cite{bib:TS,bib:C-T}
\begin{align}
\label{eq:X0X1X2Z}
 X_0 = \frac 1 Z ,\ 
 X_1= \frac \lambda Z ,\ 
 X_2 = \frac{a \lambda^2}{Z} ,\ 
 Z= 1 + \lambda + a \lambda^2 . 
\end{align}
 The fugacity $\lambda $ is specified by $ \lambda \frac{\mathrm{d}}{\mathrm{d}\lambda}\ln Z =\rho $ as 
\begin{align}\label{eq:lambda=}
 \lambda = 2\rho / ( 1-\rho+R), \ R= \sqrt{1-(1-4a)(2-\rho)\rho }.
\end{align}
 Substituting the formulas \eqref{eq:X0X1X2Z}, \eqref{eq:lambda=} into \eqref{eq:Jto}, one finds~\cite{bib:Kan} 
\begin{align} 
\label{eq:Jto:k2}
 J_\to = p_{10} \rho (2-\rho) \big[ 1-\rho + b (\rho + R -1 ) /2 \big] \big/ ( 1+R ) . \ 
\end{align}
 The current $ J = (1-q) J_\to $ depends on the jump rates, see Fig.~\ref{fig:fundamental-diagrams}. We shall occasionally use notations like $ X_m (\rho) $ and $ J_\to (\rho) $ to emphasize that they are functions of $ \rho $. 
 
In this work, we study the $k=2$ model particularly in the case where particles are extremely misanthrope, i.e., $a=0$, by performing Monte Carlo simulations. (A similar extreme case of the Katz-Lebowitz-Spohn (KLS) model \cite{bib:KLS} was analyzed in \cite{bib:Kra}.) Under the periodic boundary condition and at sufficiently large times, the probabilities of finding configurations are governed by the same master equations as the usual SEP\footnote{For $ \rho > 1 $, the dynamics (e.g. the mean-squared displacement) of an individual particle is different from the SEP.}.
When the global density $ \rho < 1$, all the sites are either occupied by one particle or empty; 
 $( X_0, X_1, X_2 ) = (1-\rho,\rho,0)$. If the target site is occupied, any particle cannot jump due to $ p_{11} = 0 $. For $\rho> 1$, no empty site appears, and we regard $\tau_i=2$ as $1 $ and $ \tau_i=1$ as $0$. Therefore $( X_0, X_1, X_2 ) = (0,2-\rho,\rho-1) $. For $\rho=1$, all the sites are occupied by one particle. In summary, the fundamental diagram (Fig.~\ref{fig:fundamental-diagrams}~(b)) of this extreme case consists of the two parabolas 
\begin{align}
\label{eq:two-parabolas}
 J (\rho) = 
 \begin{cases}
 (1-q) p_{10}\rho(1-\rho) & (0<\rho\le 1) , \\
 (1-q) p_{21} ( \rho- 1) (2-\rho) & (1<\rho< 2) .
 \end{cases}
\end{align}

\section{Open boundary conditions}

\begin{figure} 
\vspace{-3mm}\begin{center}
  \includegraphics[width=80mm]{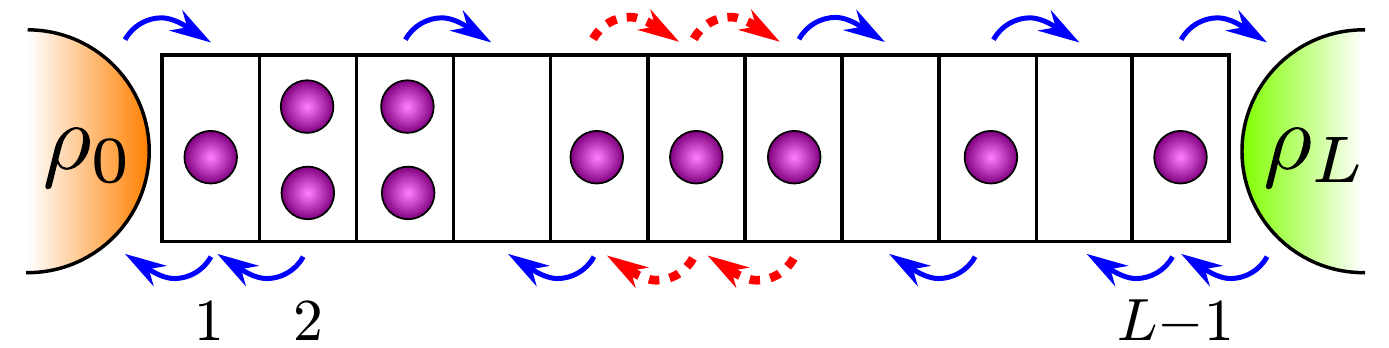}
 \caption{ Misanthrope process with $k=2$, 
 connected to density reservoirs at both ends. The arrows represent possible jumps of particles. In the extreme case ($ p_{11}=0 $), the dashed lines correspond to prohibited jumps. 
 \label{fig:schematic}} 
\end{center}
\end{figure} 

 Let us denote the occupation number at site $ i $ by $ \tau_i \in \{ 0, 1, \dots, k\} $ for the general misanthrope process with general $k$. We consider the situation, where a finite chain with $L-1$ sites is connected to two density reservoirs (Fig.~\ref{fig:schematic}). At the left (right) boundary, particles are injected and extracted with rates $ \alpha_{\tau_1} $ ($ \delta_{\tau_{L-1}} $) and $ \gamma_{\tau_1} $ ($ \beta_{\tau_{L-1}} $), respectively, depending on $ \tau_1 $ ($ \tau_{L-1} $). To realize the reservoir densities $ \rho_0 $ and $ \rho_L $ of virtual sites $ i=0,L $, we set 
\begin{align} 
\label{eq:leftboundary}
& \alpha_\tau = \sum_{ m \ge 1 } p_{m\tau} X_m (\rho_0 ), \ 
 \beta_\tau = \sum_{ m \ge 0 } p_{\tau m} X_m (\rho_L ),\\ 
& \gamma_\tau = q \sum_{ m \ge 0 } p_{\tau m} X_m (\rho_0 ), \ 
 \delta_\tau = q \sum_{ m \ge 1 } p_{m\tau} X_m (\rho_L ) . 
\label{eq:rightboundary}
\end{align} 
 Note that, in general, the product measure does not give the correct stationary state, except for the case $ \rho_0=\rho_L $. 

The case $(k,q) = (1,1)$ is the symmetric SEP, where the boundary rates \eqref{eq:leftboundary}, \eqref{eq:rightboundary} satisfy $ \alpha_0 + \gamma_1 = \beta_1 + \delta_0 = p_{10} $, and the current and the density profile are given by \cite{bib:D,bib:S} 
\begin{align}
 J ( \rho_0 , \rho_{L} ) & = p_{10} ( \rho_0 - \rho_{L} ) /L , \\ 
 \rho_i & = \rho_0 + ( \rho_L - \rho_0 ) i / L ,
 \label{eq:single-lane-J-rho}
\end{align}
in the stationary state. Therefore the diffusivity is simply given by the jump rate $p_{10}$. The correlation function $ C =\langle \tau_i \tau_{i+1} \rangle - \langle \tau_i \rangle \langle \tau_{i+1} \rangle $ is also exactly calculated as \cite{bib:D,bib:S}
\begin{align}
 C & = - i (L-i-1) (\rho_0-\rho_L)^2 / [L^2(L-1) ] \\ 
 & \simeq - x(1-x) (\rho_0-\rho_L)^2 / L \quad (x= i/L,\ L\to \infty ) .
 \label{eq:correalation-SSEP}
\end{align}

The extreme misanthrope process $ (k,p_{11}) =(2,0 ) $ with open boundaries has an equivalence to the SEP, which is similar to what we explained in the previous section. In the cases $ \rho_0 < 1 \wedge \rho_L < 1 $ and $ \rho_0> 1 \wedge \rho_L> 1 $, physical quantities, such as the stationary density profile and current, are essentially the same as the SEP. Therefore we shall show simulations only for the case $ \rho_0 > 1 > \rho_L $. 

\vspace{-1mm}
\section{Symmetric case} 
 In this section we consider the symmetric misanthrope process $ (k,q) = (2,1) $. With Kronecker's delta $ g_m (\tau) = 1 (\text{for } m=\tau), 0 (\text{for }m\neq \tau )$, the current between sites $i$ and $i+1$ is expressed as 
\begin{align}
\label{eq:Ji=<G>}
 J_i &= \langle G(\tau_i,\tau_{i+1} ) \rangle \quad(0<i<L-1), \\
\label{eq:J0=<Gt>}
J_0 &= \langle \tilde G (\rho_0,\tau_1) \rangle ,\quad 
J_L = - \langle \tilde G ( \rho_L ,\tau_{L-1} ) \rangle , \\
 G(\tau,\sigma) 
 &= \sum_{m=1,2 \atop n=0,1} p_{mn} \Big( g_m(\tau) g_n(\sigma) - g_n(\tau) g_m(\sigma) \Big) , 
 \label{eq:f(tau,sigma)} \\
\tilde G (\rho,\tau) 
&= \sum_{m=1,2 \atop n=0,1} p_{ m n } \Big( X_m (\rho ) g_n (\tau ) - X_n ( \rho ) g_m (\tau ) \Big) . 
\end{align}
Using polynomial representations $ g_0(\tau) = g_2 (2-\tau) = \tfrac{ (1-\tau) (2-\tau) }{2},\ g_1(\tau) = \tau(2-\tau) $, one can transform \eqref{eq:Ji=<G>} and \eqref{eq:J0=<Gt>} into the gradient form $ J_i = K_i - K_{i+1} $, where
\begin{align}
\label{eq:Ki=}
K_i = & 
 \begin{cases}
 p_{10}\langle g_1(\tau_i) \rangle +p_{20}\langle g_2(\tau_i) \rangle & ( 0<i<L ), \\ 
 h (\rho_i) & (i=0,L ) , 
 \end{cases} \\ 
 \label{eq:hrho=}
 h ( \rho) : =& p_{10} X_1 (\rho ) +p_{20} X_2 (\rho ) \\ 
 =& p_{10 } \rho \big[ 2-\rho + b (\rho+R-1) /2 \big] \big/ (1+R) . 
 \end{align}
This property leads to the stationary current in the form\footnote{As a byproduct, one finds 
$ K_i = h ( \rho_0 ) + \big[h ( \rho_L ) - h ( \rho_0 )\big] i / L $. In the case $ 2 p_{10} = p_{20} $ ($\Leftrightarrow p_{10 } = p_{21} \Leftrightarrow b=2$), $ K_i $ is identical to $ p_{10} \rho_i $, shown by Eqns. \eqref{eq:Ki=}, \eqref{eq:hrho=}, hence this formula is interpreted as the linear density profile $ \rho_i = \rho_0 + ( \rho_L - \rho_0) i/ L $. }
\begin{align}
 & J_0 = J_1 = \cdots = J_{L-1} = 
 \frac{1}{L} \sum_{i=0}^{L-1} ( K_i - K_{i+1} ) \\ 
 & = \big[ h ( \rho_0 ) - h ( \rho_L ) \big] / L
 =: J ( \rho_0 , \rho_L ) .
 \label{eq:J=} 
\end{align}
 We assume Fick's law in the coarse-grained view $ x = i /L $ 
\begin{align}
\label{eq:Ficks}
 D ( \rho (x) ) \frac{\mathrm{d}}{ \mathrm{d}x } \rho (x) = - L J ( \rho_0 , \rho_L ) . \end{align}
 Integrating both sides with respect to $x$ in the interval $[ 0,1 ] $, one gets $ \int_{ \rho_0 }^{\rho_L} D ( \rho ) \mathrm{d} \rho = - L J ( \rho_0 , \rho_L ) $. The diffusivity should be given as
 $ D(\rho) = \frac{ \mathrm{d} }{ \mathrm{d} \rho } h ( \rho ) $, 
 and more explicitly\footnote{This formula can be also expressed as
 $D(\rho) = \chi^{-1} J_\to ( \rho )$ with 
 $\chi = \sum_{m=0,1,2} X_m(\rho) \big( m - \rho \big)^2= \lambda \frac{\mathrm{d}\rho}{\mathrm{d}\lambda} $
 and the rightward current \eqref{eq:Jto:k2}.
One may derive this formula by assuming the product measure with the weights \eqref{eq:X0X1X2Z} in the bulk sites \cite{bib:B}. In our case we did not use this assumption to achieve \eqref{eq:D=}.}
\begin{align} 
\label{eq:D=}
 D(\rho) = p_{10} \big[ 1 - \rho + b(\rho + R -1 )/2 \big] / R . 
\end{align} 
Integrating again both sides of \eqref{eq:Ficks} with respect to $x$ in the interval 
$ [ 0, x ] $, the density profile is implicitly found as 
\begin{align}
\label{eq:prediction}
& h (\rho(x)) - h ( \rho_0 ) = - x L J ( \rho_0 , \rho_L ) . 
\end{align}
 Though the formula for the current \eqref{eq:J=} is correct for any finite $L$, the prediction \eqref{eq:prediction} does not always provide the true analytic formula. However, we expect that the density profile $ \langle \tau_{ xL } \rangle $ converges to \eqref{eq:prediction} in the limit $ L\to \infty $.

\begin{figure} 
\vspace{-3mm}\begin{center}
  \includegraphics[width=40mm]{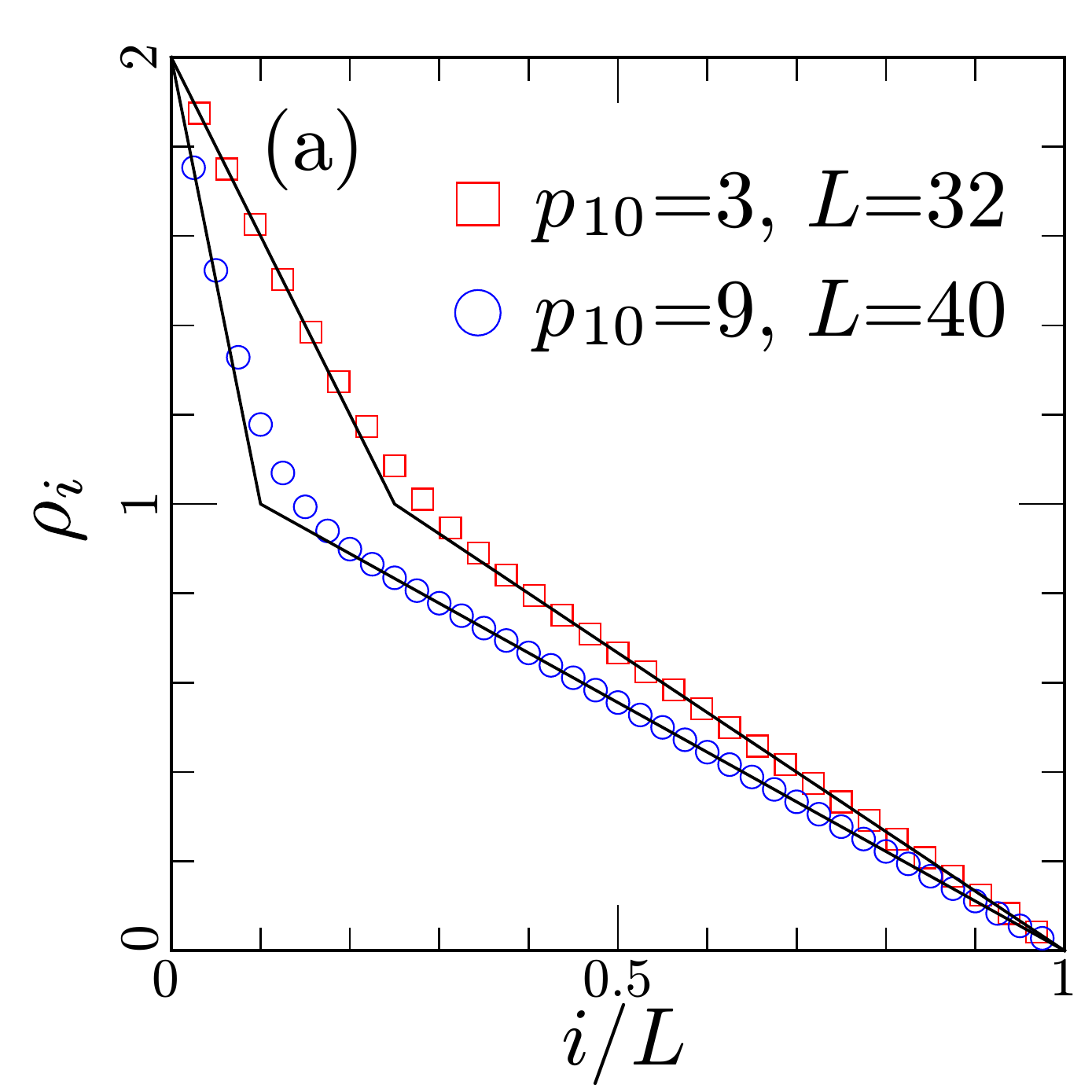}
  \includegraphics[width=40mm]{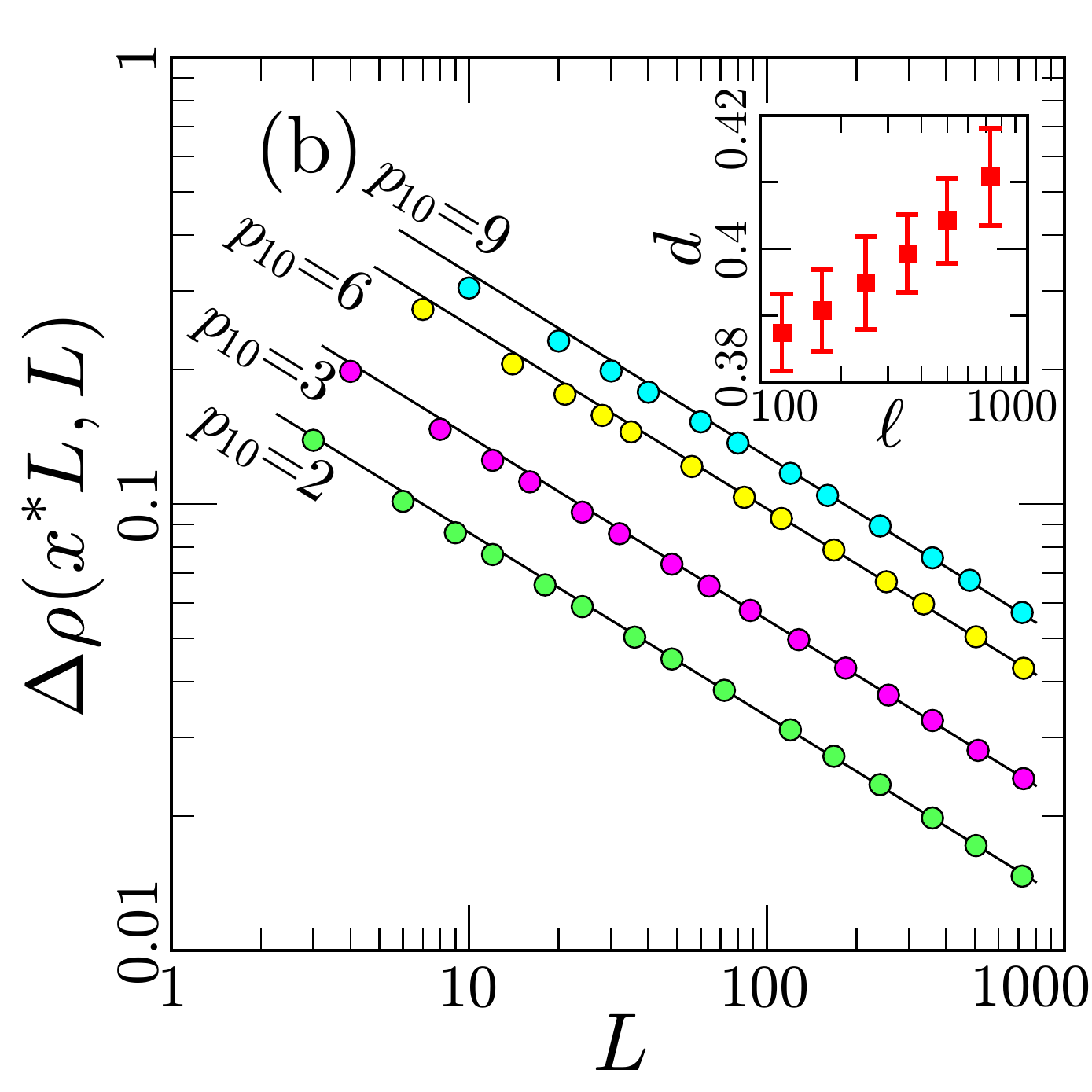}
  \includegraphics[width=40mm]{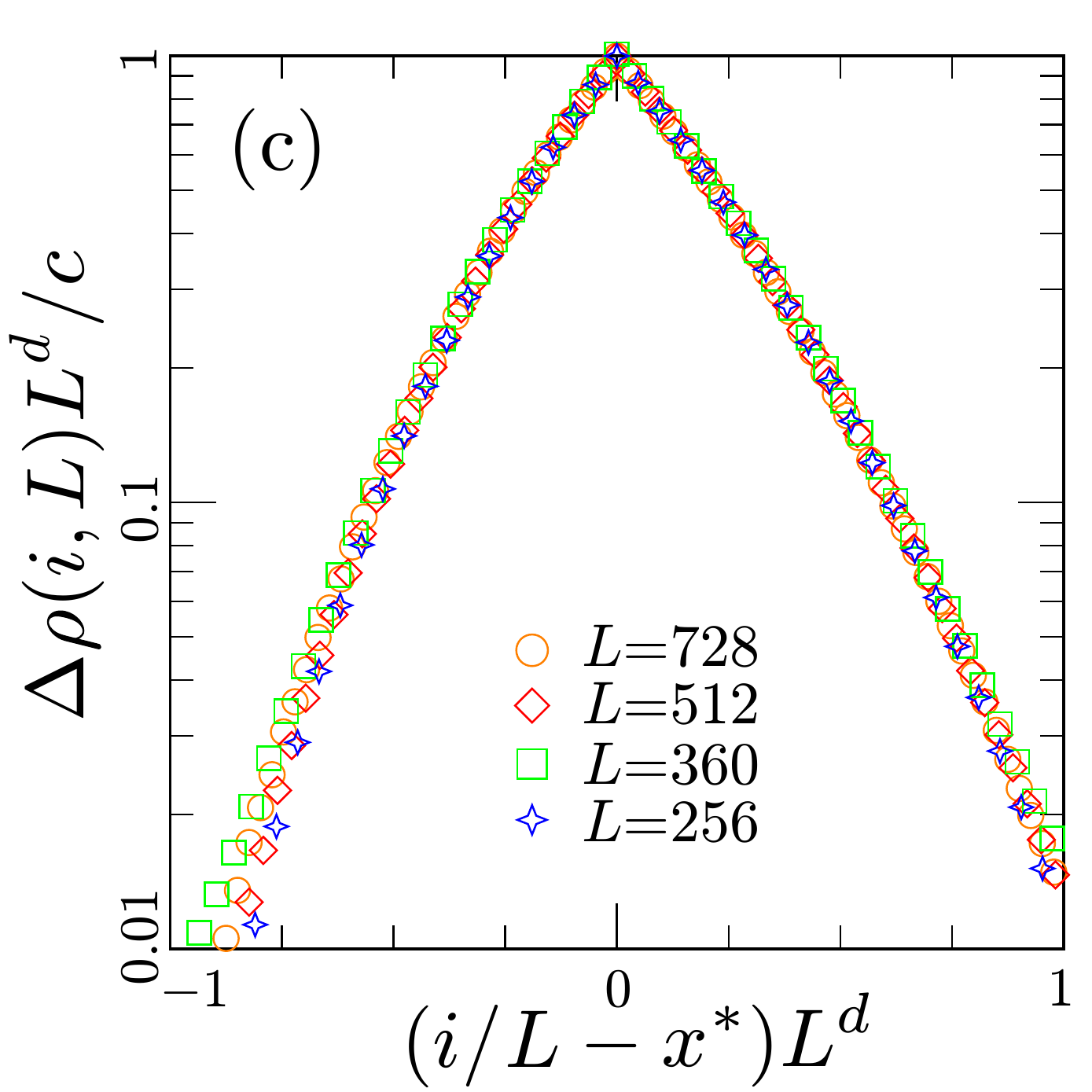}
  \includegraphics[width=40mm]{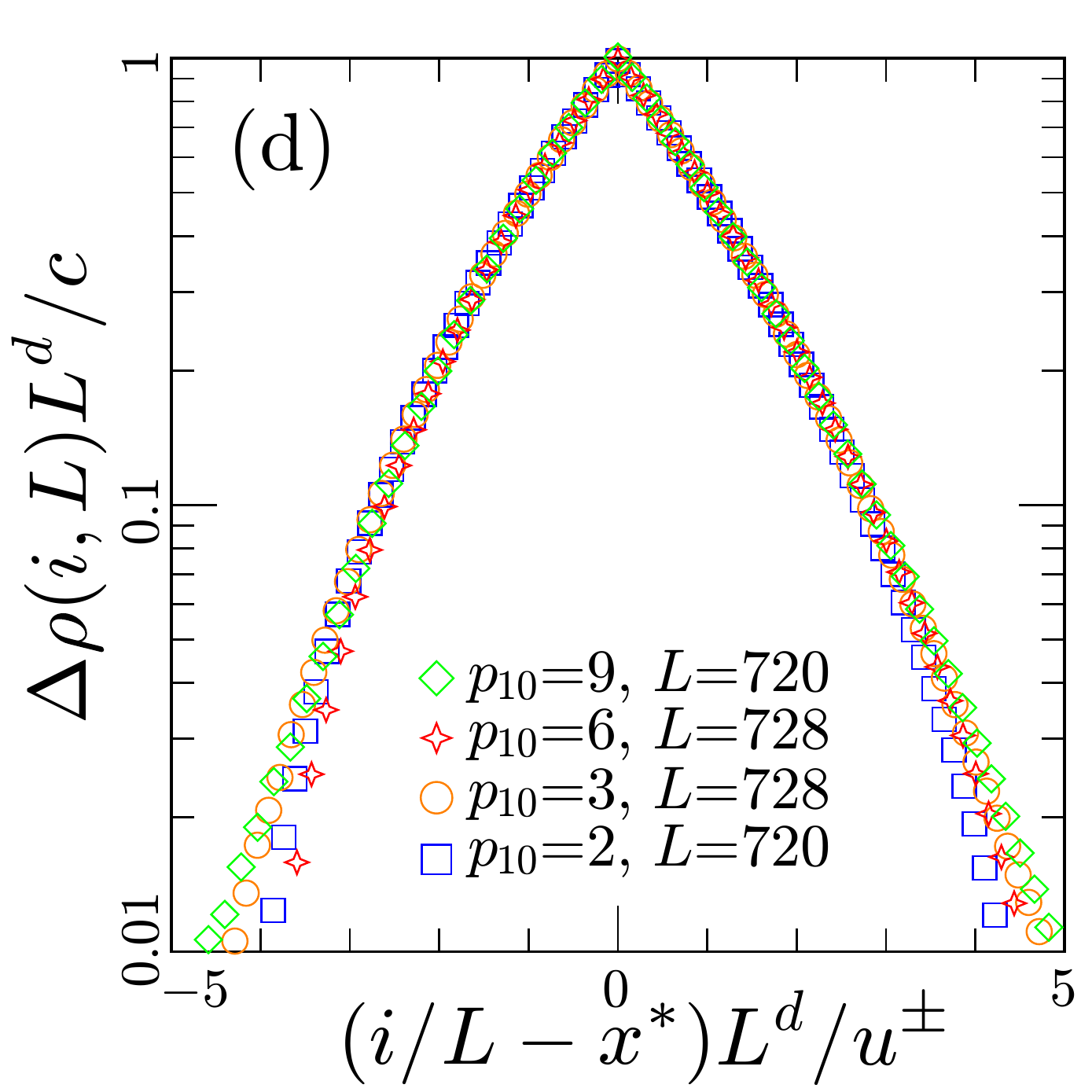}
\caption{ (a) Density profiles of the symmetric misanthrope process with
 $p_{21} = 1$. The values of $ p_{10} $ and $ L $ are given in the panel. 
 (b) The discrepancies at site $ i =x^* L $ for various values of $p_{10}$.
 The simulation results and numerical fittings are indicated by circles and solid lines, respectively. 
 (c) Rescaled discrepancies vs. rescaled site for various values of $L$ with $ (p_{10}, p_{21} ) = (3,1) $. 
 (d) Rescaled discrepancies vs. rescaled site for various values of $p_{10}$. 
 For all the four panels we have set the boundary densities as $ ( \rho_0,\rho_L ) = (2,0)$. 
 For each set of the parameters $ ( p_{10} ,L ) $ we performed two independent simulation runs. 
 We also averaged the simulation data over $ 10^6 \le t \le 10^9 $ for $ p_{10}=2,3 $ 
 or $ 10^6 \le t \le 10^8 $ for $ p_{10}=6,9 $. 
 \label{fig:symmetric-density-profiles}} 
\end{center}
\end{figure}

 Now we turn to the extreme case $ ( k,q, p_{11} ) = ( 2,1,0 ) $ with $ \rho_0 > 1> \rho_L $. The diffusivity \eqref{eq:D=} reduces to 
\begin{align}\label{eq:D=p}
 D(\rho) = p_{10}\, (\text{for}\ \rho < 1 ) , \ p_{21}\, (\text{for}\ \rho > 1 ) , 
\end{align} 
 and the prediction \eqref{eq:prediction} becomes piecewise linear 
\begin{align}
 \rho (x) =& \begin{cases} \rho_0 + \frac{x}{x^*} (1-\rho_0) & (0<x<x^* ), \\
 \rho_L + \frac{1-x}{1-x^*} (1-\rho_L) & (x^*<x<1 ) , \end{cases} \\ 
 x^* =& ( \rho_0 - 1 ) p_{21} / \big[ ( 1 - \rho_L ) p_{10} + ( \rho_0 - 1 ) p_{21} \big] . 
\end{align}
The position $x^*$ separates the space into high- and low-density domains, i.e., $ \rho (x) > 1 $ for $ x<x^* $ and $ \rho (x) < 1 $ for $ x>x^* $. We observe that the simulation results are deviated from the prediction near $x=x^*$, see Fig.~\ref{fig:symmetric-density-profiles}~(a).

Let us denote by $ \Delta \rho (i,L) = \rho_i - \rho(i/L) $ the difference between the true density $ \rho_i $ measured by simulations and the prediction. Figure~\ref{fig:symmetric-density-profiles}~(b) shows the discrepancy at the site $ i = L x^* $ vs. $L$. It seems that it exhibits power-law decay $ \Delta \rho ( x^*L , L ) \sim c (b) L^{-d} $, and we assume that the exponent $d$ is independent of $b$. The inset shows our numerical estimation of $d$, depending on the interval of the system size $ \ell /5 < L < \ell $ that we used, e.g. $ d\approx 0.411 $ for $ \ell \approx 720 $. (The error bars are due to changing the value of $p_{10}$.) For the fitting lines in (b) and rescaling of the x- and y-axes of (c) and (d), we use this value\footnote{We think that the power-law decay with the exponent slightly bigger than $0.4$ is the most reasonable guess. However the following possibility has not been excluded: the slopes in the logarithmic frame continue to slowly decrease and diverge in the limit $ L \to \infty$.}. We observe overlap of markers for different values of $L$ in Fig.~\ref{fig:symmetric-density-profiles}~(c), $ L^d \Delta \rho ( x L , L ) /c$ vs. $ ( x - x^* ) L^d $. Furthermore we can find \textit{rescalers} $ u^+(b) >0 $ and $ u^-(b) < 0 $, such that the discrepancies of different values of $b $ also show overlap, see Fig.~\ref{fig:symmetric-density-profiles}~(d).
Technically we determined $ u^{\pm} (b) $ via 
 $ x^{\pm}$ $( x^- < x^* < x^+ ) $ as 
\begin{align} 
 \Delta \rho ( x^{\pm} L ,L ) L^d &= c (b) /2 , \ 
 u^{\pm} (b) = ( x^{\pm} - x^* ) L^d .
\end{align} 
 This result indicates the existence of a scaling function $ f $ 
\begin{align} 
 \Delta \rho (xL,L) \stackrel{ L\to \infty}{\simeq} 
 \begin{cases} 
 c(b) L^{-d} f \big( ( u^+(b) )^{-1} (x-x^*) L^d \big) & (x\ge x^*) ,\\ 
 c(b) L^{-d} f \big( - ( u^-(b) )^{-1} (x-x^*) L^d \big) & (x<x^*) . 
 \end{cases}
\end{align}

\begin{figure} 
\vspace{-3mm}\begin{center}
  \includegraphics[width=40mm]{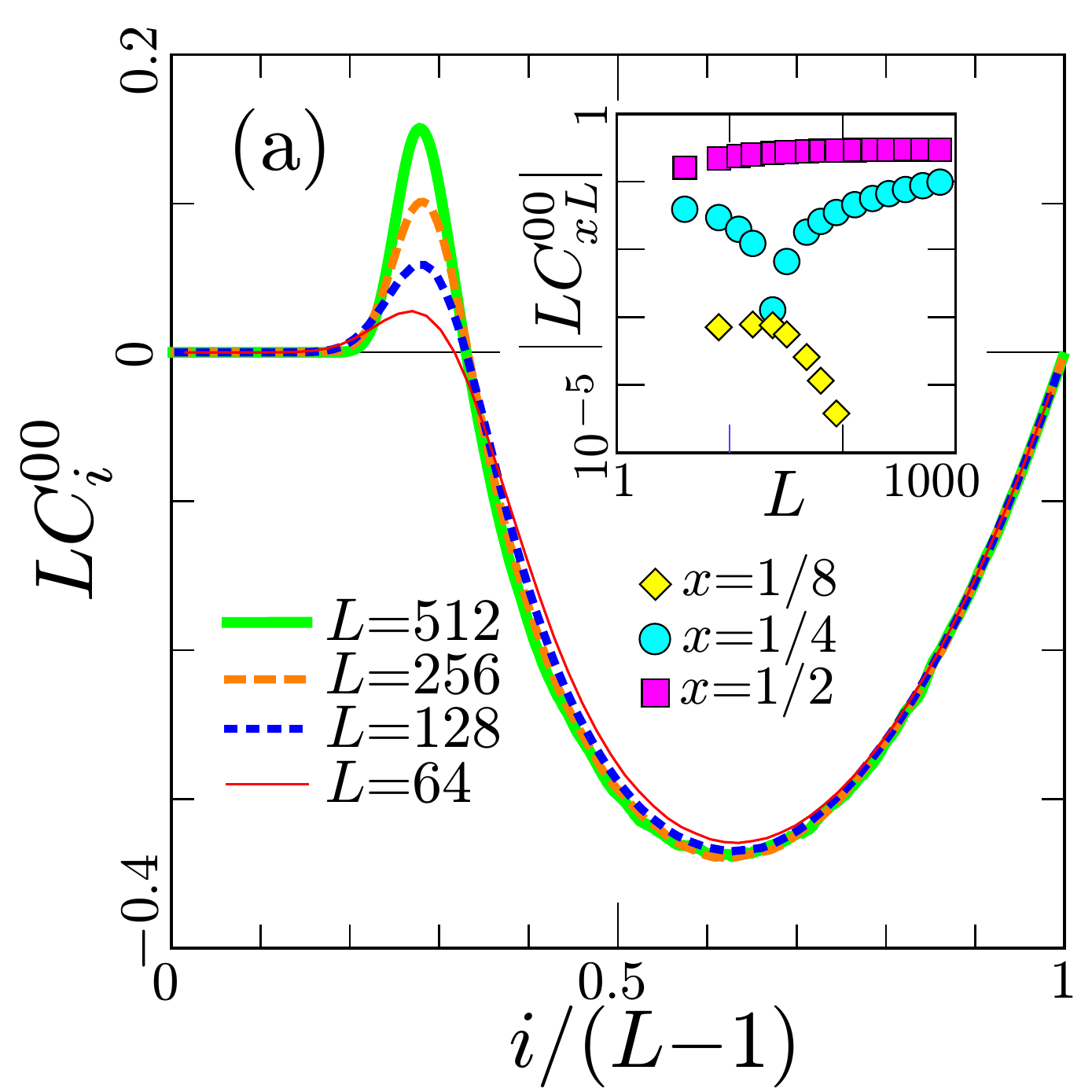}
  \includegraphics[width=40mm]{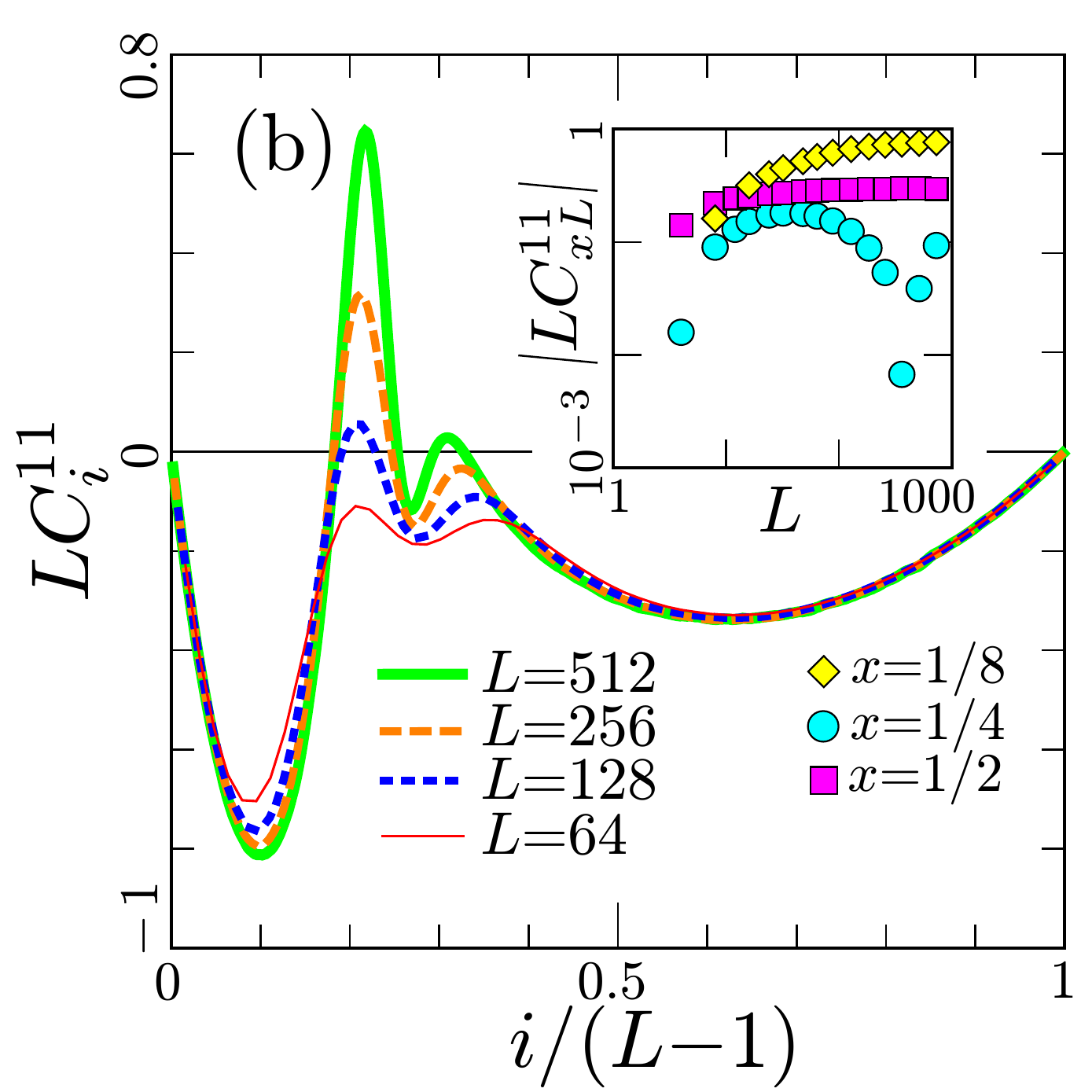}
  \includegraphics[width=40mm]{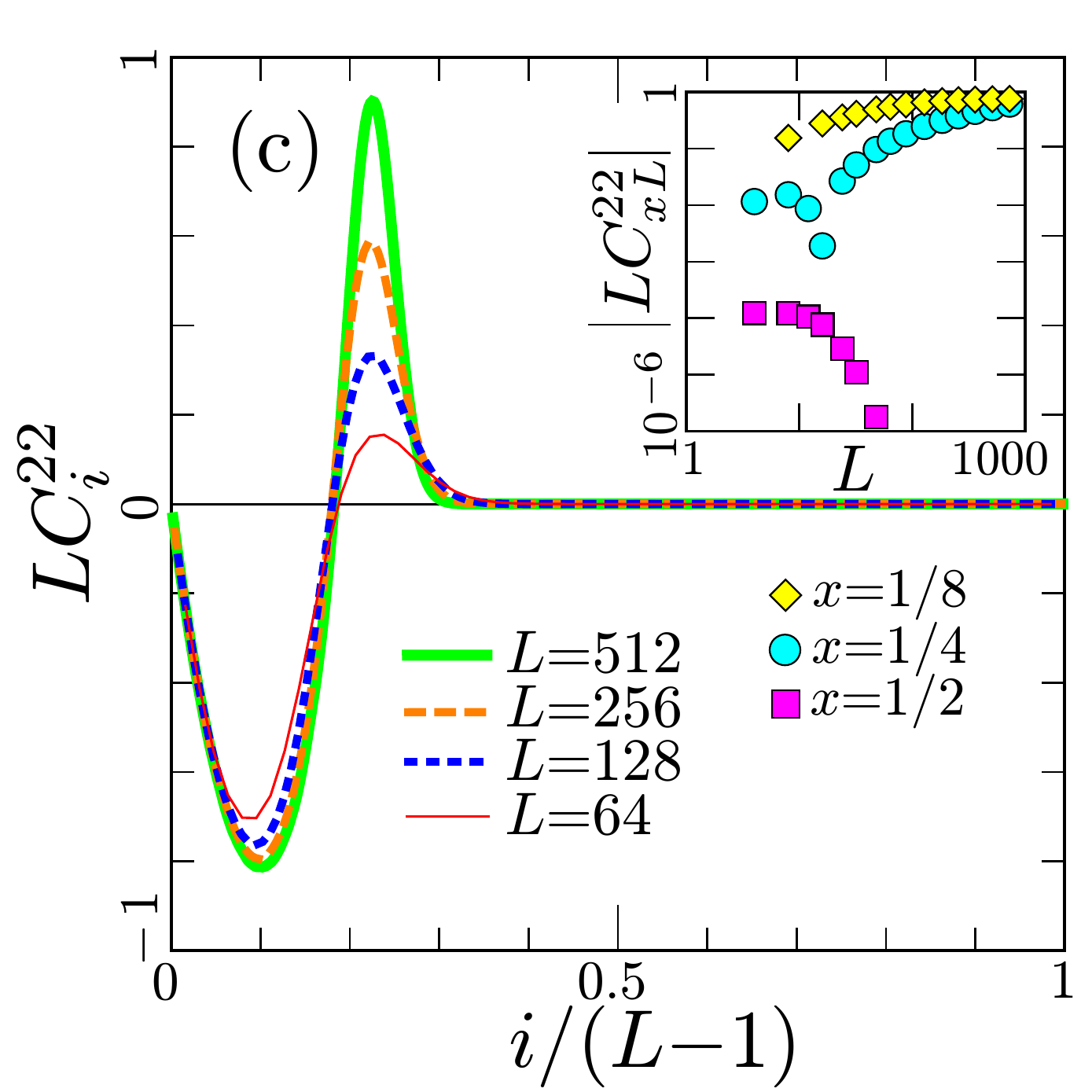}
  \includegraphics[width=40mm]{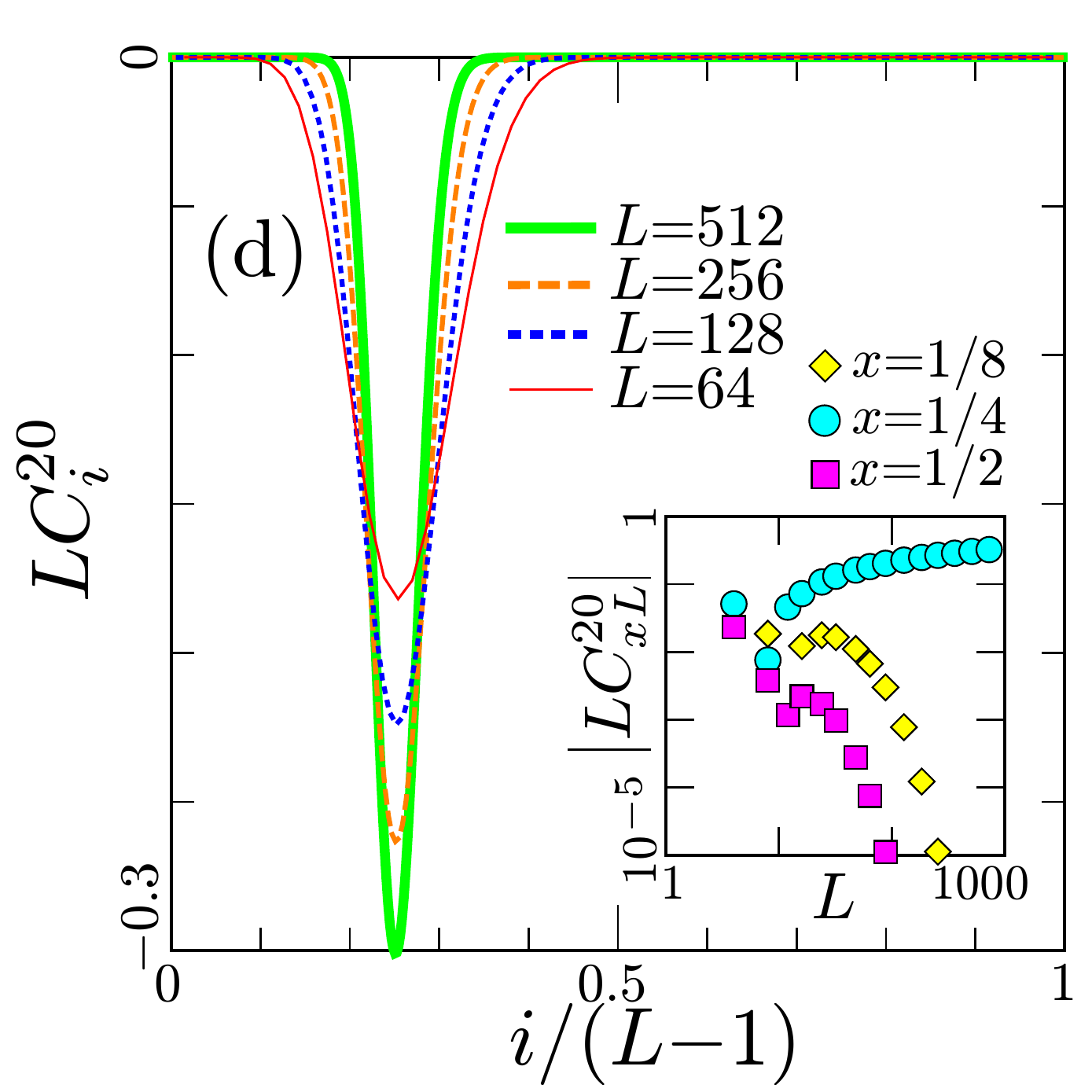}
 \caption{ Correlation profiles. 
 The parameters were chosen as 
 $ ( p_{10} , p_{21} , \rho_0 , \rho_L ) = (3,1,2,0)$ ($ x^*= 1/4 $).
 For each $L$, we performed two independent simulation runs. 
 We also averaged the simulation data over $ 10^6 \le t \le 10^9 $. 
 \label{fig:correlation-profiles}}
\end{center}
\end{figure}

Now we turn to nearest-neighbor correlation functions $ C_i^{ m n } = \langle g_m (\tau_i) g_n (\tau_{i+1}) \rangle - \langle g_m (\tau_i)\rangle \langle g_n (\tau_{i+1}) \rangle $ 
 for $m,n\in \{0,1,2\}$, see Fig.~\ref{fig:correlation-profiles} for simulation results. It seems that, far from $ i/(L-1) \simeq x = x^*$, $ C_i^{ m n }$'s decay being proportional to $L^{-1} $ as the SEP \eqref{eq:correalation-SSEP} or faster than the power law, since the lines of different values of $L $ are overlapping or almost 0. On the other hand, in the vicinity of $x = x^* $, they decay more slowly than $ L^{-1} $. The insets also support these observations. (Another scenario could be that the finite-size effects are too strong to observe $ O(L^{-1}) $-decay).

\section{Totally asymmetric case}

\begin{figure} 
\vspace{-3mm}\begin{center}
   \includegraphics[width=40mm]{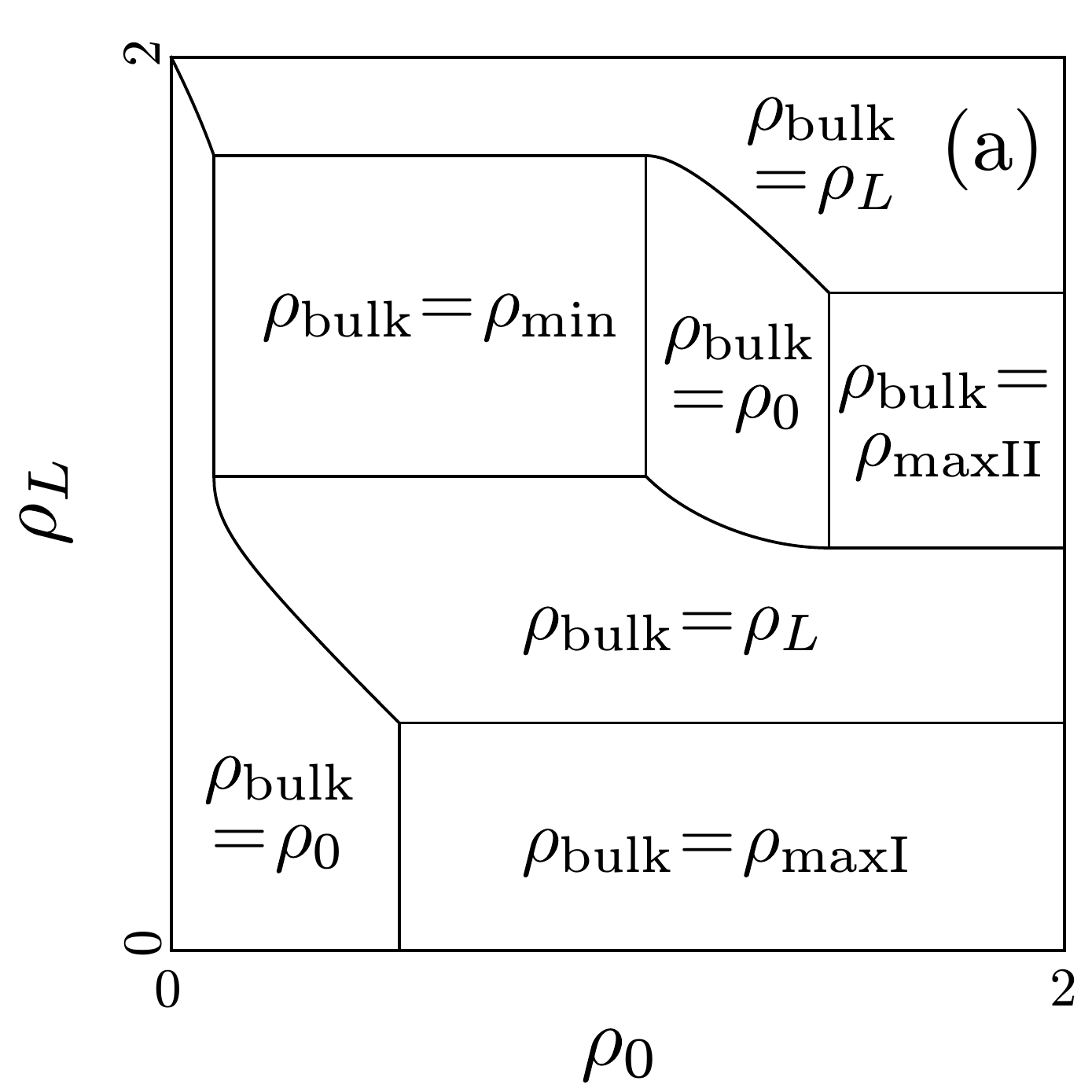}
   \includegraphics[width=40mm]{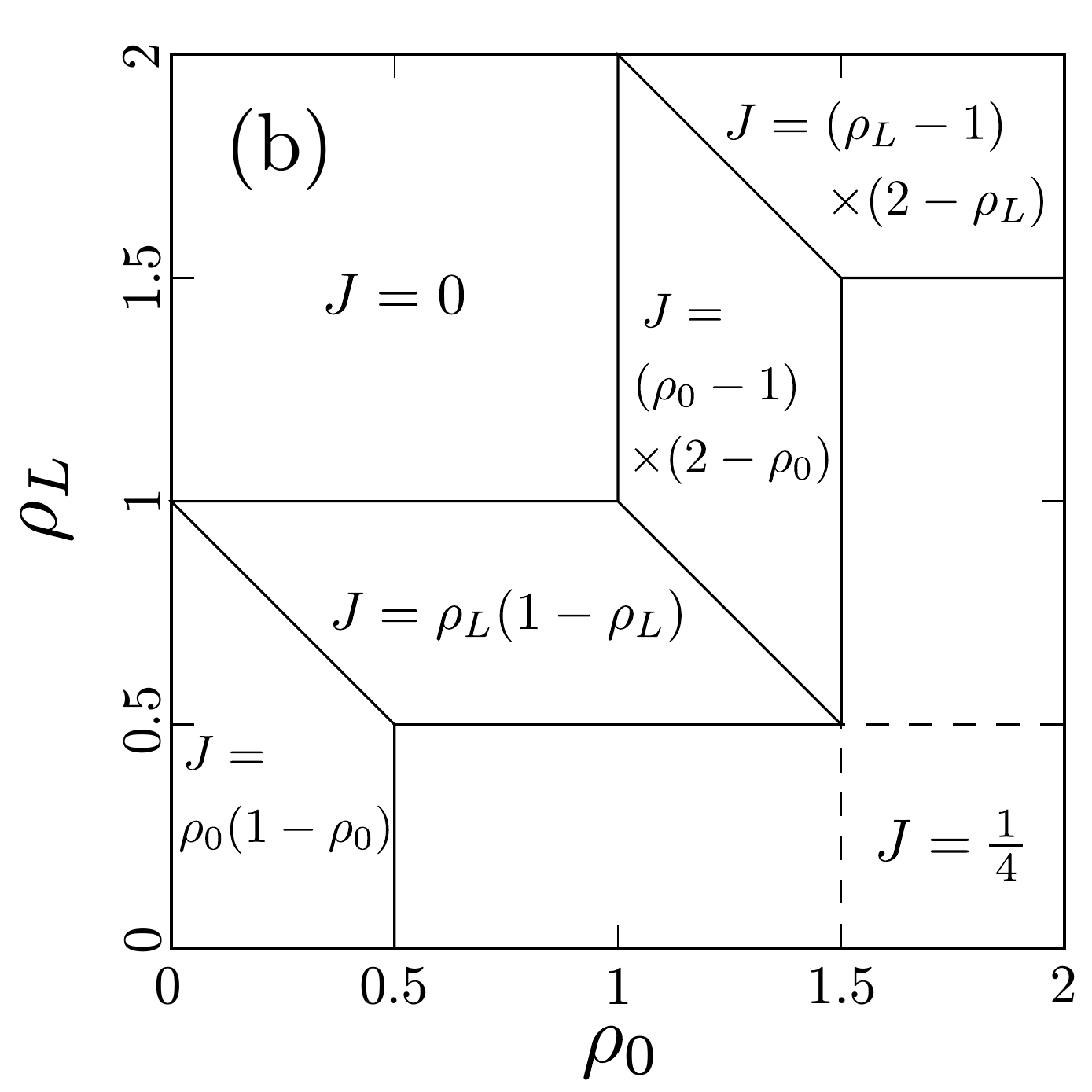}
\caption{ 
(a) Phase diagram of the totally asymmetric misanthrope process 
 in the case where $J(\rho)$ has two local maxima
 $ J(\rho_\text{maxI}) > J(\rho_\text{maxII}) $
 $ ( \rho_\text{maxI}<\rho_\text{maxII} ) $. 
 The stationary current is give by $ J( \rho_\text{bulk} ) $,
 in particular, $J(\rho_\text{min})$ is the local minimum. 
 (b) Phase diagram for $p_{11} = 0 \wedge p_{10} = p_{21}=1 (a=0,b=2)$. 
\label{fig:phase-diagrams}} 
\end{center}
\end{figure}

We consider the totally asymmetric case $ (k,q ) = (2,0)$. Since we know the fundamental diagram even with $p_{11}$ arbitrary (Fig.~\ref{fig:fundamental-diagrams}), the phase diagram (Fig.~\ref{fig:phase-diagrams} (a)) is predictable by the extremal current principal. See the original works \cite{bib:PS,bib:HKPS} for details e.g. the phase boundaries. In addition to the extreme misanthropy $p_{11} = 0$, we set $ p_{10} = p_{21} =1$, where the two maximal currents have the same value. The phase diagram of the current is simplified as Fig.~\ref{fig:phase-diagrams} (b).

 We wish to explore properties of an ``anti-shock''~\cite{bib:BS} appearing in the case $ \rho_0 > 1 > \rho_L $. Instead of the ``second-class particle'' \cite{bib:BCFG} used in the SEP, we introduce another microscopic definition of the shock position $S$. Denote by $S_2$ the rightmost site occupied by two particles, and by $S_0$ the leftmost empty site. Any configuration is written as 
 \begin{align} \tau_1 \cdots \tau_{S_2-1} 2 11 \cdots 11 0 \tau_{S_0+1} \cdots \tau_{L-1} \end{align}
with $ \tau_i \in \{ 1,2 \} $ for $ i<S_2 $ and $ \tau_i \in \{ 0,1 \} $ for $ i>S_0 $. In particular, $ S_2=0$ when there is no site occupied by two particles, and $ S_0 = L $ when there is no empty site. Then we simply set $ S = \frac{S_2 + S_0}{2} \in \{ \frac 1 2 , 1 , \frac 3 2, \cdots , \frac{2L-1}{2} \} $. (See \cite{bib:CHA-R,bib:dGF} for similar microscopic definitions of shocks.) Note that we always have $ S_2 < S_0 $ due to $ \rho_0 > 1 > \rho_L $. The \textit{tag} $S_2 $ increases by one, according to jump of a particle at site $ S_2 $ if $ \tau_{S_2+1} = 1$. On the other hand, the tag $S_0 $ decreases by one, according to jump of a particle at site $ S_0 -1 $ if $ \tau_{S_0-1} = 1$. These events shift the shock position rightward or leftward by $1/2 $. When $ S_0 - S_2 = 1 $, a particle at site $S_2$ jumps to site $S_0 $, which causes a \textit{non-local} shift; in this situation, the tags are renewed, i.e. the second rightmost doubly occupied site becomes the new $S_2$ and the second leftmost empty site becomes the new $S_0$. The configuration of Fig.~\ref{fig:schematic} is an example; if a particle on the 3rd site jumps to the 4th, the shock position changes as $ ( S_2, S, S_0 ) = ( 3, 7/2 , 4 ) \to (2, 5 , 8) $. 
 
\begin{figure}
\begin{center}
  \includegraphics[width=40mm]{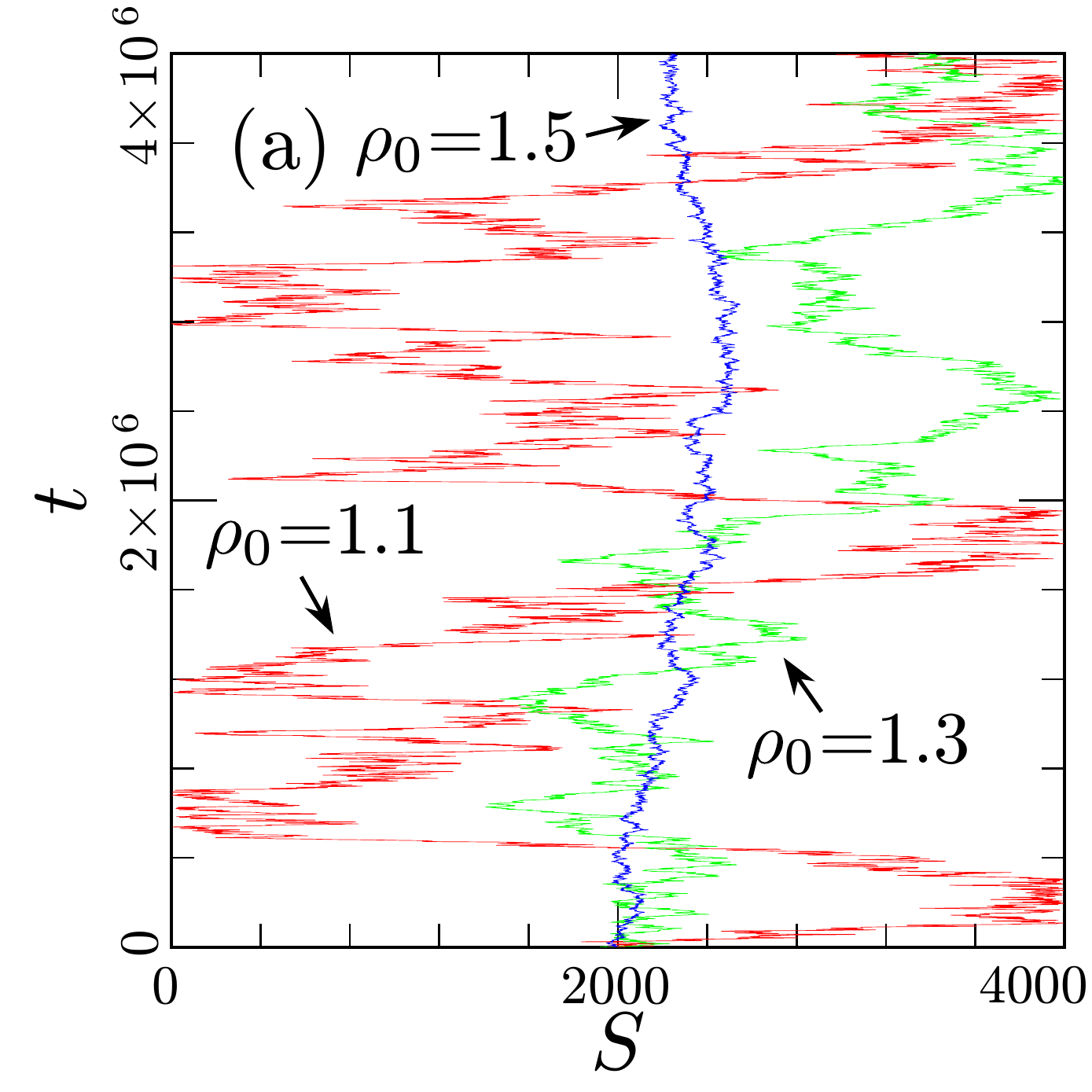}
  \includegraphics[width=40mm]{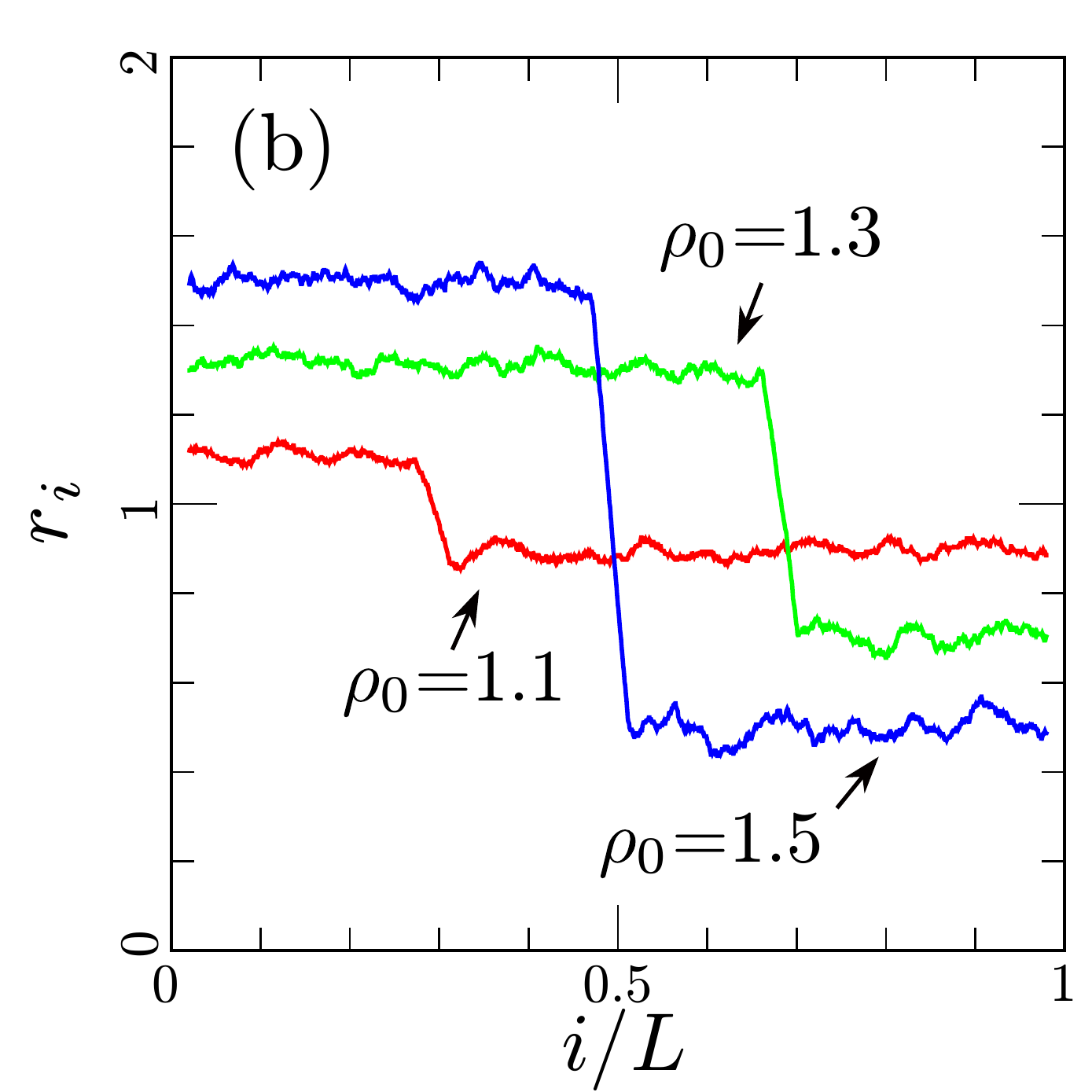}
  \includegraphics[width=40mm]{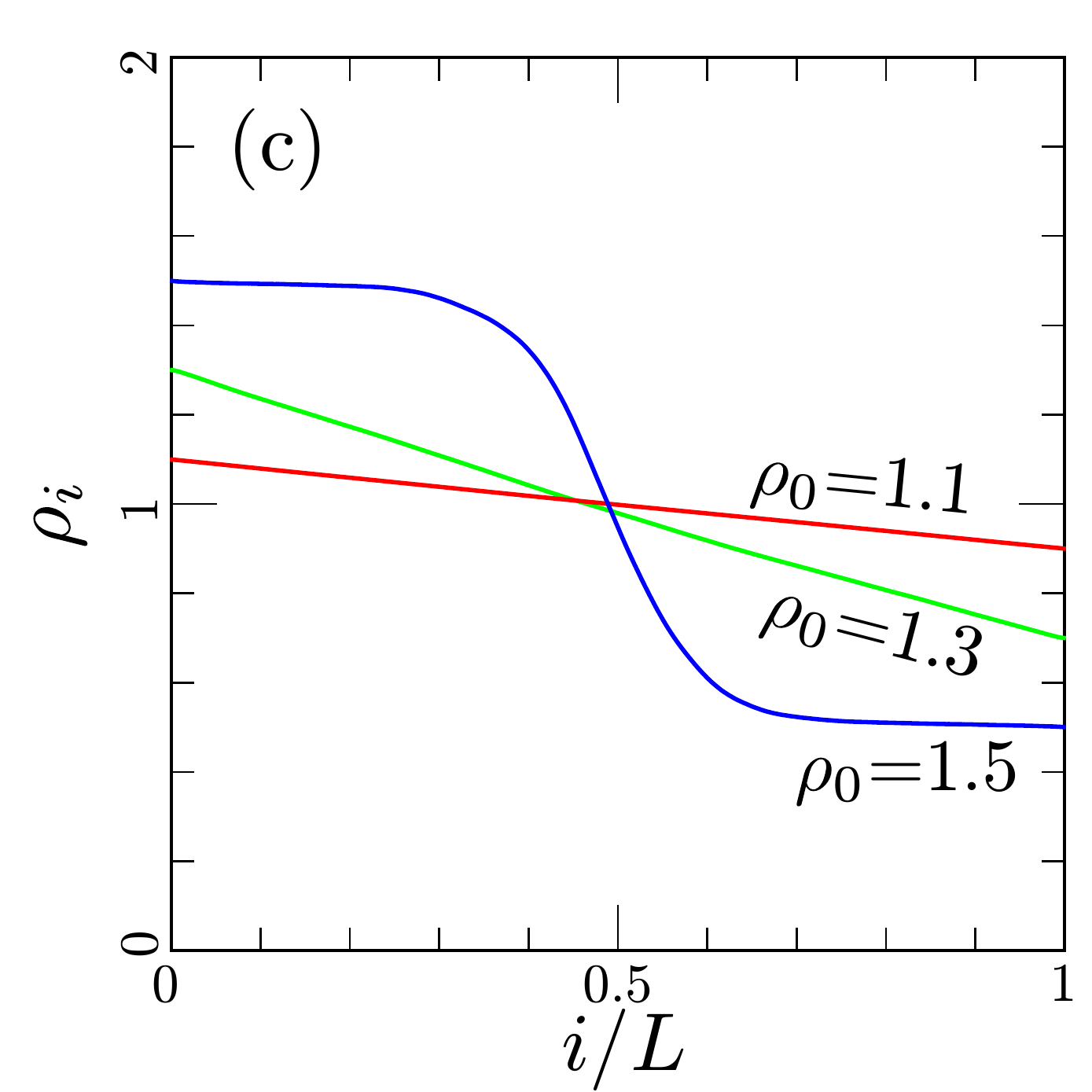}
  \includegraphics[width=40mm]{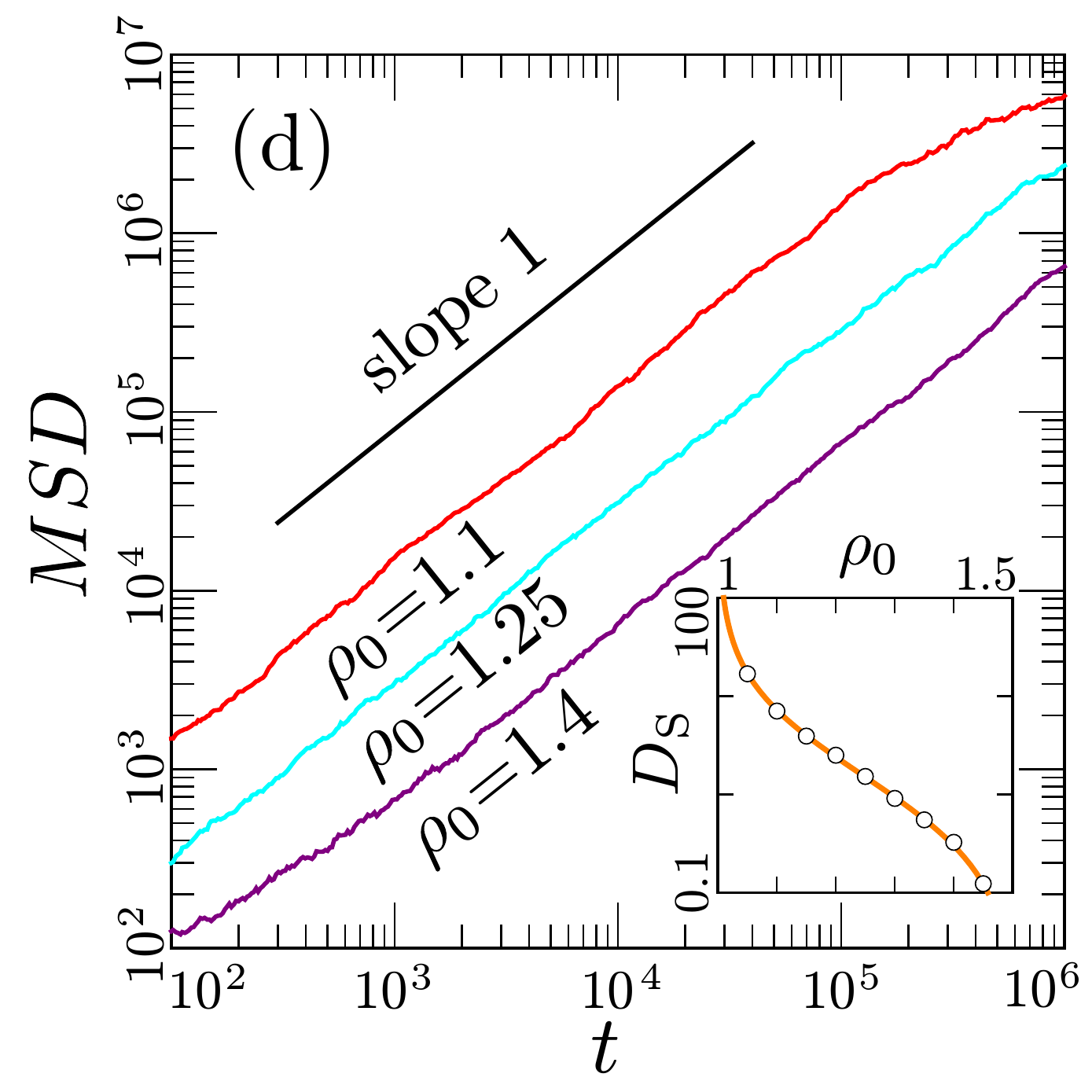}
\caption{ 
Simulation results on the phase transition line $1 < \rho_0 = 2 - \rho_L < 1.5$. 
(a) Kymograph of shocks in single simulation runs with $L=4000$ and the initial position $ s= 0.5 $. 
(b) Density profiles $ r_i $ for $ L=10^4 $, made from snapshots. 
(c) Density profiles $ \rho_i $ for $L=10^3$, averaged over time interval $ 10^6 \le t\le 10^8 $ 
 and over two independent runs. 
(d) Mean-squared displacement vs. time and (inset) diffusivity $ D_\text{S} $, 
 by averaging over 500 simulation runs. 
\label{fig:diffusion}} 
\end{center}
\end{figure}

The shock is delocalized on the phase transition line $1 < \rho_0 = 2 - \rho_L < 1.5$, as shown in Fig.~\ref{fig:diffusion} (a). The densities of the domains $ i <S $ and $ i>S $ are given by the reservoir densities $ \rho_0 $ and $ \rho_L $, respectively \cite{bib:HKPS}. Figure~\ref{fig:diffusion} (b) shows typical density profiles of spatial average $ r_i : = \frac{1}{2\ell+1}\sum_{k=-\ell}^\ell \tau_{k+i} $ with $ \ell = 100 $, by using snapshots $ \tau_1\cdots \tau_{L-1} $ without time or ensemble average. Under the assumption that the motion of the shock is governed by a random walk with reflective boundaries, the density profile averaged over long time becomes linear, connecting $ \rho_0 $ and $\rho_L$, see Fig.~\ref{fig:diffusion} (c). As shown in Fig.~\ref{fig:diffusion} (d), there is a regime where the mean-squared displacement, $ MSD= \langle ( S(t+t_0) - S(t_0) )^2 \rangle $, is proportional to time $t$, supporting the random-walk description. In the inset, we plot the diffusivity $D_\text{S} $, which was estimated from simulation data $ \frac{MSD}{2t} $ in $t \in [ T/100,T ] $. We chose a value of $T $ so as to avoid the saturation of the linearity. The line is a guessed form $D_\text{S} = \frac{ (2-\rho_0 ) ( 3-2\rho_0 ) }{ 4(\rho_0-1) } $, which approaches 0 as $ \rho_0 \to 1.5 $. Actually, on the point $ \rho_0 = 2- \rho_L = 1.5 $, the location of the shock is restricted to the vicinity of $ S=L/2 $, see Fig.~\ref{fig:diffusion} (a), (b), and (c).

\begin{figure}
\begin{center}
  \includegraphics[width=40mm]{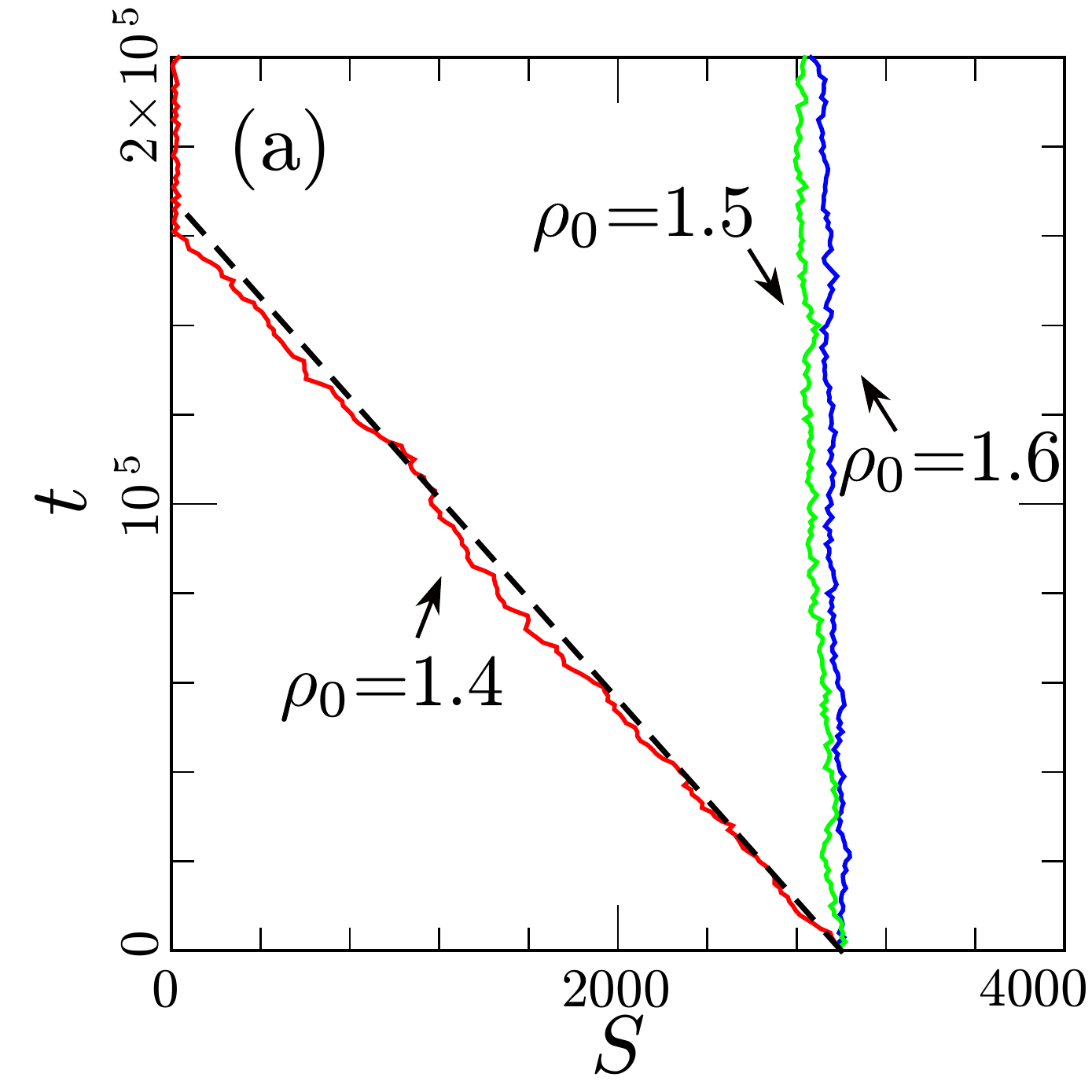}
  \includegraphics[width=40mm]{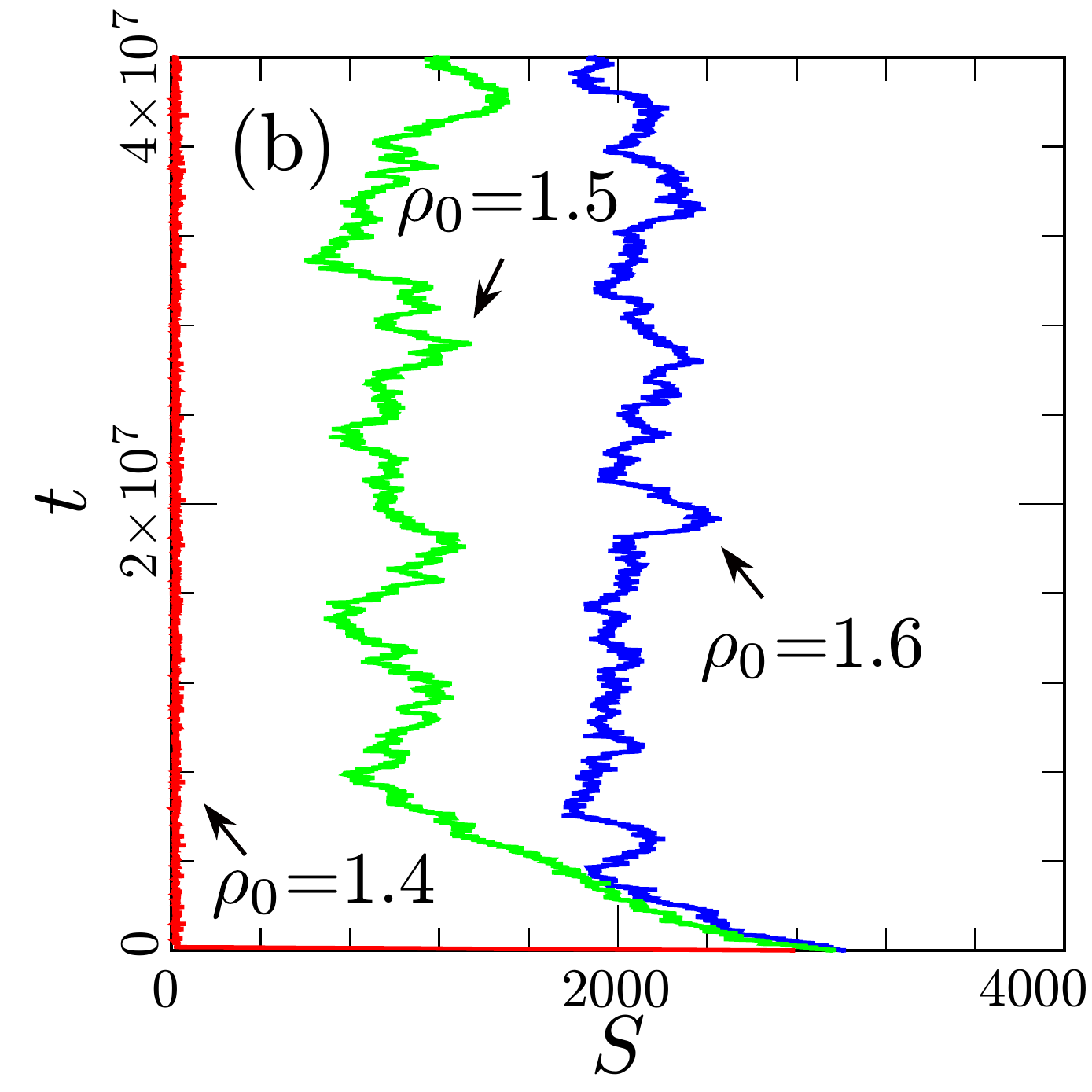}
  \includegraphics[width=40mm]{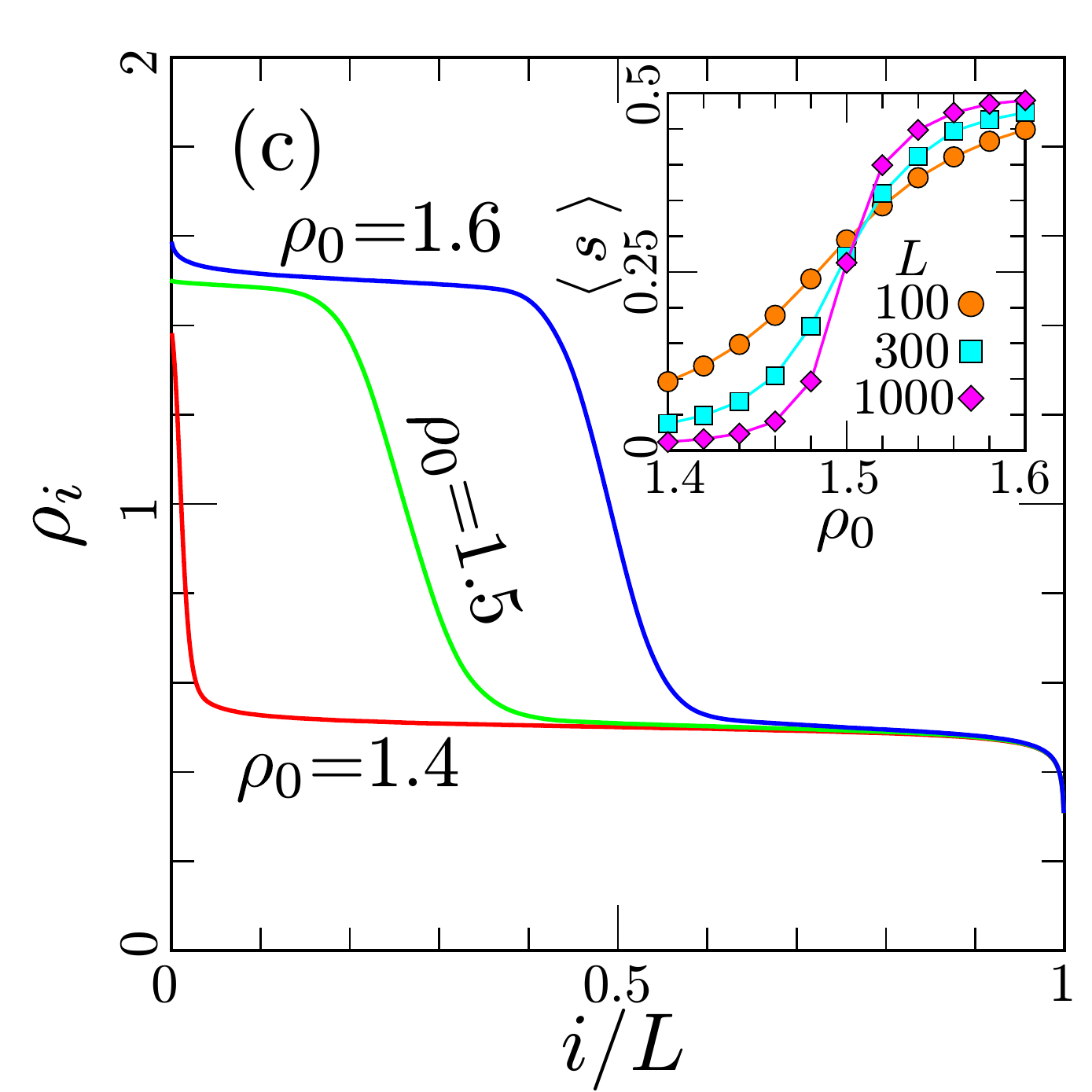}
  
  \includegraphics[width=40mm]{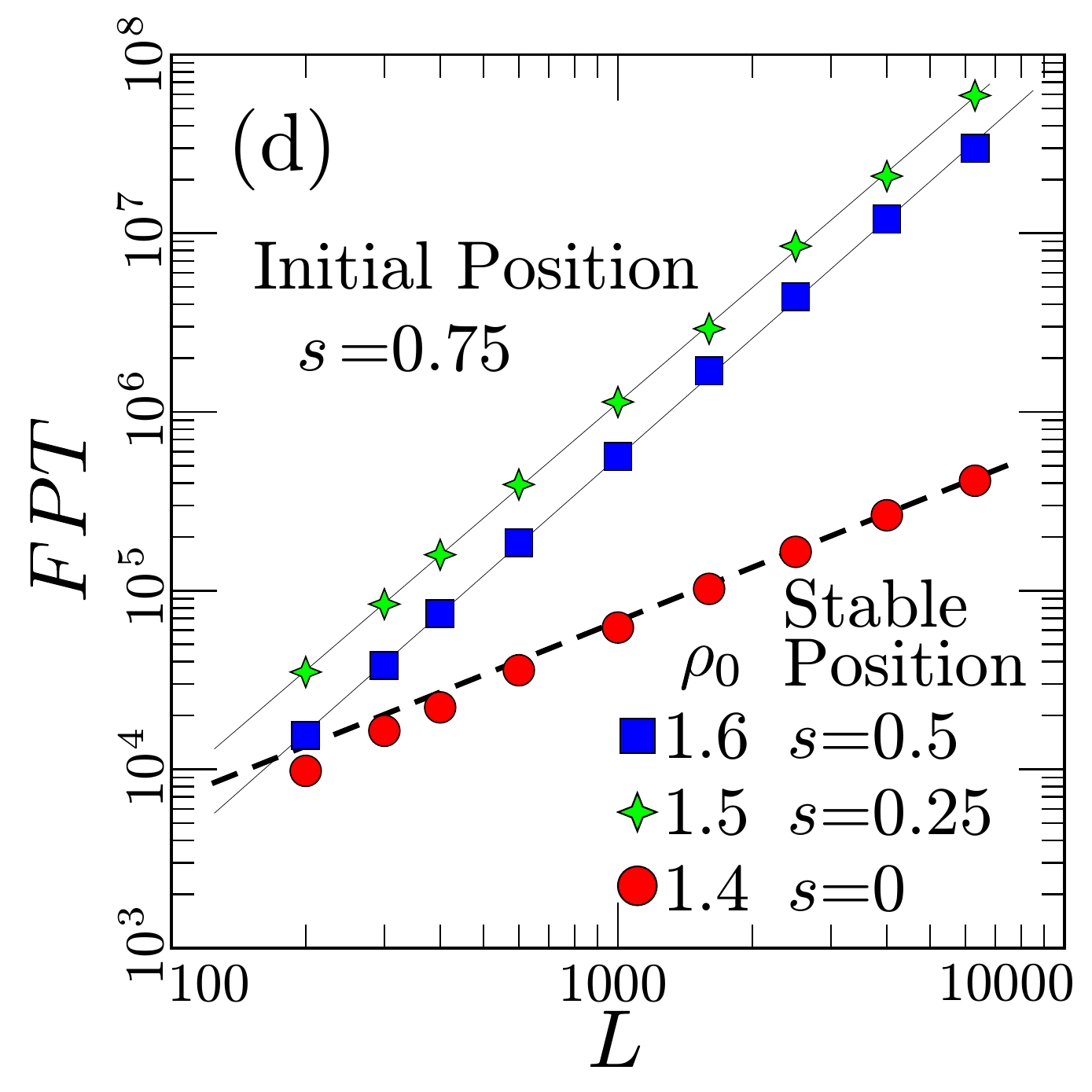}
  \includegraphics[width=40mm]{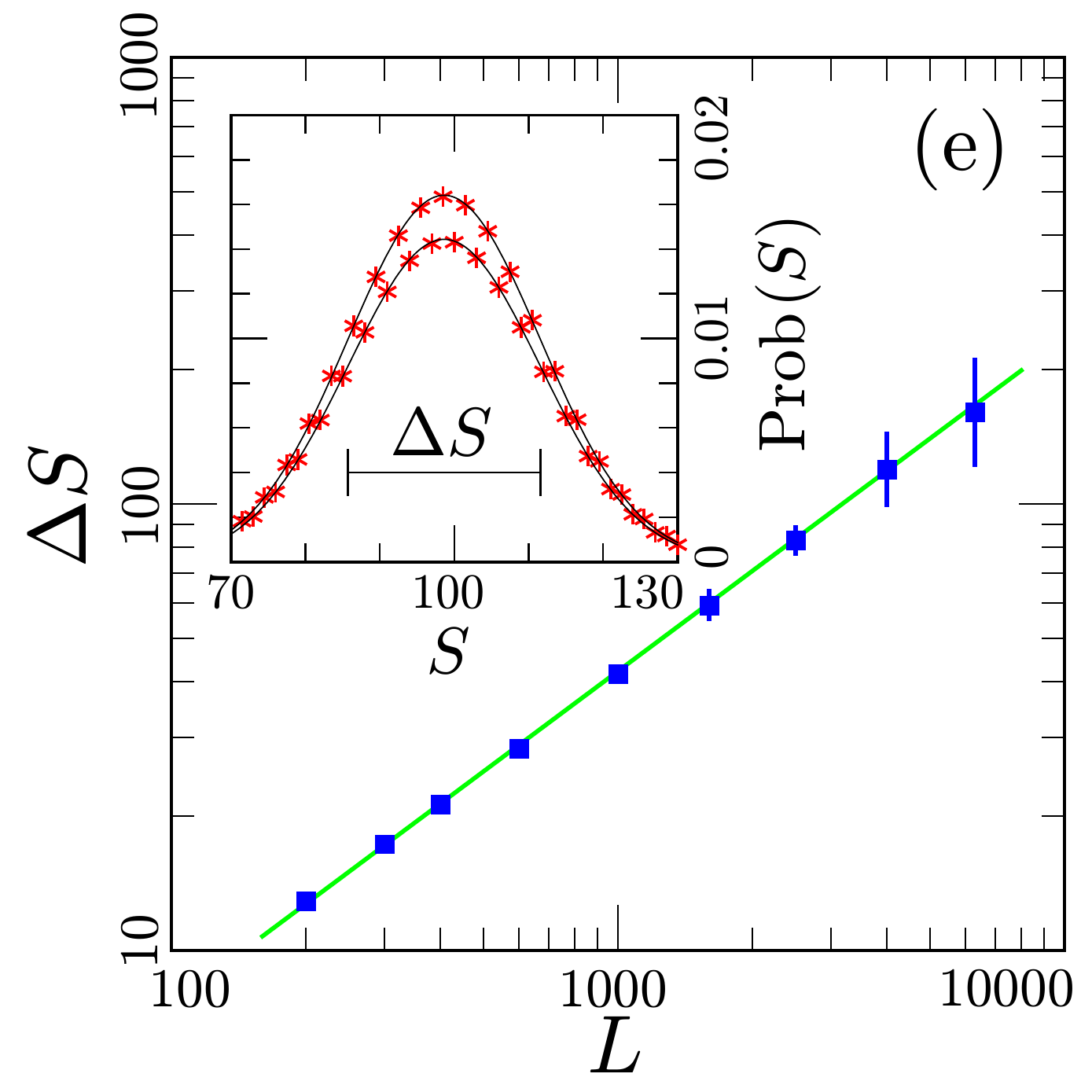}
  \includegraphics[width=40mm]{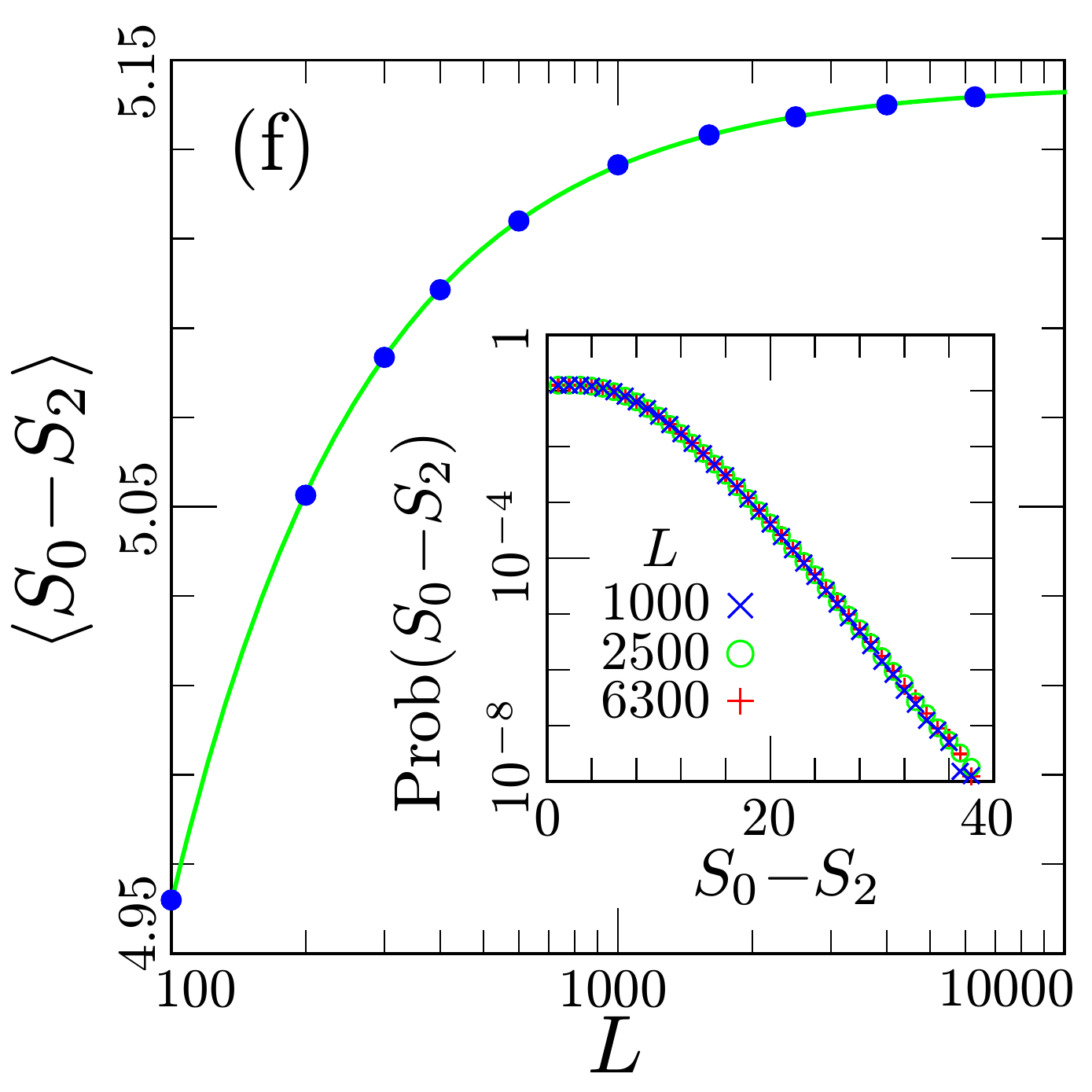}
\caption{ Simulation results in the maximal current phase with $ \rho_L = 0.2 $. 
(a,b) Kymographs of the shock in single simulation runs with 
 $ L=4000$ and the initial position $ S=0.75L $. 
 The dashed line is the theoretical line $ S = t v_\text{S} + 0.75L $. 
(c) Density profiles 
 for $ L=10^3 $,
 and (inset) mean shock positions near the sub-phase boundary with lines as a guide to the eyes.
(d) First passage time. The dashed line is $FPT = - 0.75L/v_\text{S} $, and thin lines are fitting. 
(e) Shock width 
 for $ (\rho_0, \rho_L) = (1.7,0.2) $,
 and (inset) probability distribution of the shock position for $L=200$. 
(f) Mean instantaneous shock width 
 for $ (\rho_0, \rho_L) = (1.7,0.2) $,
 and (inset) probability distribution. 
 The solid line is a fitting curve in the form $ v-wL^{-1} $. 
 Technical details: We performed two independent runs for (c), $ 10^5/L $ runs for (d) or ten runs for (e) and (f). 
 We also averaged over $ 10^6 \le t\le 10^8 $ for (c), (e) and (f). 
\label{fig:max-current}} 
\end{center}
\vspace{-3mm}
\end{figure}

Now we examine properties of the shock in the maximal current phase. We denote the rescaled shock position by $ s = S /L $. In the sub-phase $ \rho_0 < 1.5 \wedge \rho_L < 0.5 $, the left and right domain densities are $ ( \rho_\text{Left} , \rho_\text{Right} ) = ( \rho_0 ,0.5) $ from the general framework \cite{bib:PS,bib:HKPS}. The shock velocity is given by $ v_\text{S} = \frac{ J( \rho_\text{Left} ) - J ( \rho_\text{Right} ) }{ \rho_\text{Left} -\rho_\text{Right} }< 0 $ with $ J (\rho) $ Eqn.~\eqref{eq:two-parabolas}. The shock position linearly achieves the vicinity of the left reservoir, as we see an example $ (\rho_0,\rho_L) = (1.4, 0.2 )$ in Fig.~\ref{fig:max-current}~(a). In the sub-phase $ \rho_0 > 1.5 \wedge \rho_L < 0.5 $ and its boundaries (dashed lines in Fig.~\ref{fig:phase-diagrams}~(b)), the domain densities become $ ( \rho_\text{Left} , \rho_\text{Right} ) = (1.5,0.5) $. Therefore we have $ v_\text{S}= 0 $. Indeed, in Fig.~\ref{fig:max-current} (a), we cannot see a clear tendency of the shock motion. By changing the time scale as Fig.~\ref{fig:max-current} (b), however, we observe that the shock moves to $ \langle s\rangle = 0.5 $ in this sub-phase, and $ \langle s\rangle = 0.25 $ on $\rho_0 = 1.5\wedge \rho_L< 0.5 $. The density profiles in Fig.~\ref{fig:max-current} (c) also imply that $ \langle s\rangle = 0.5 $ and $ 0.25 $ are the \textit{stable} positions. (The localization of the shock in the middle of the system was previously shown for the repulsive KLS model in the seminal work~\cite{bib:Kru}.) In the inset, the deviations from these values are observed, which are expected to be finite-size effect. On the other sub-phase boundary $\rho_0 < 1.5\wedge \rho_L = 0.5 $, $ \langle s\rangle = 0.75 $ because of a symmetry. We also measured the first passage time $FPT$, i.e. the first time when the shock \textit{hits} a stable position, see Fig.~\ref{fig:max-current} (d). The estimated exponents $ z $ ($ FPT=O( L^z ) $) for $ \rho_0=1.5$ and $1.6 $ are $\approx2.14$ and $\approx2.21$, respectively, while $z=1$ for $ \rho_0=1.4 $ due to $ v_\text{S}\neq0 $ (we regard $ \langle s \rangle = 0 $ as a stable position).

In Fig.~\ref{fig:max-current} (e), the standard deviation of the shock position $\Delta S = \sqrt{ \langle S^2 \rangle - \langle S \rangle^2 } $ (shock width) seems to exhibit a power law $\Delta S \sim L^e$. The exponent $e$ from fitting is $ e \approx 0.74$ for $\rho_0 = 1.7$. We also performed fitting for other values of $ \rho_0 $ with $ \rho_L=0.2$, and found the exponent between $ 0.7 < e < 0.8 $ (not shown here). This result is different from 1/2 and 1/3 observed in the exclusion process with a single-site defect \cite{bib:JL}. The probability distribution of $S$ (inset) consists of two curves, according to whether $ S $ is an integer or a half-integer. Both of them are well fitted by Gaussian distributions. Another shock width $ S_0 - S_2 $ is instantaneously measured in a given configuration. Figure~\ref{fig:max-current} (f) implies that its average converges to some value as $ L\to \infty $, which is different from $ \Delta S$. We expect that the decay of its probability distribution is asymptotically exponential, see the inset of Fig.~\ref{fig:max-current} (f).

\section{Conclusions} 

We investigated the $k=2$ misanthrope process with open boundaries, in particular we showed simulation results in the extreme case. 

For the symmetric case ($q=1$), the true density profile is, in general, deviated from the hydrodynamic prediction, where the point $ x^*$ separates between high and low density domains. We expect that the discrepancy exhibits a power-law decay at $ x=x^* $ and there exists a scaling function in the vicinity of $ x=x^* $. The nearest-neighbor correlations decay more slowly than $ O(L^{-1}) $ for $x \approx x^* $. 
 
For the totally asymmetric case ($q=0$), an anti-shock can exist, and we introduced a microscopic definition of its position. On the transition line $ \rho_0 + \rho_L =2 $ ($ 1 < \rho_0<1.5 $), where the motion of the shock position is diffusive, we probed the mean-squared displacement and measured the diffusivity characterizing the motion of the shock. In the sub-phase $ \rho_0 > 1.5 \wedge \rho_L< 0.5 $ of the maximal current phase at its boundaries, the formula of the shock velocity becomes 0. However we found that the shock very slowly reaches a stable position. The exponent of the shock width was found between 0.7 and 0.8.

\section*{Acknowledgements} 
CM is supported by CREST, JST and JSPS Grant-in-Aid No.~15K20939.

\end{document}